\documentclass{PoS}

\usepackage[outdir=./]{epstopdf}
\usepackage{nicefrac}

\title{Testing the Standard Model under the weight of heavy flavors}

\ShortTitle{Heavy Flavors}

\author{C.M.~Bouchard\thanks{current address:  The College of William and Mary, Williamsburg, Virginia}\\
        The Ohio State University, Columbus, Ohio\\
        E-mail: \email{bouchard.chris.m@gmail.com}}

\abstract{
I review recently completed (since Lattice 2013) and ongoing lattice calculations in charm and bottom flavor physics.
A comparison of the precision of lattice and experiment is made using both current experimental results and projected experimental precision in 2020.
The combination of experiment and theory reveals several tensions between nature and the Standard Model.
These tensions are reviewed in light of recent lattice results.
}

\FullConference{The 32nd International Symposium on Lattice Field Theory,\\
		23-28 June, 2014\\
		Columbia University New York, NY}

\begin{document}

\section{Introduction}
\label{sec-intro}
This review outlines recent (since last year's lattice conference) lattice QCD activity in heavy flavor physics and compares the precision of lattice and experiment for heavy flavor observables.
This comparison is made using current experimental results and projections for the year 2020.\footnote{2020 projections are based on expectations for the LHCb upgrade with $20\ {\rm fb}^{-1}$, BESIII with $20\ {\rm fb}^{-1}$, and BelleII with $50\ {\rm ab}^{-1}$.}
The combination of theory and experiment reveals several tensions between the Standard Model (SM) and nature.
These tensions are reviewed in light of recent lattice results.
After motivating lattice efforts in heavy flavor physics in the present section,
section~\ref{sec-CKM} focuses on processes that occur at tree-level in the SM, useful in precision determinations of CKM matrix elements.  
It is subdivided into leptonic (section~\ref{sec-leptonic}) and semileptonic (section~\ref{sec-semileptonic}) decays.
Section~\ref{sec-rare} discusses processes that occur via flavor changing neutral currents (FCNCs), i.e. rare decays (section~\ref{sec-decays}) and neutral meson mixing (section~\ref{sec-mixing}).
SM contributions to rare processes are small and there is a possibility of discernible new physics effects.
Each subsection is further subdivided into $D$ and $B$ processes.
 
\begin{figure}[t!]
{\scalebox{1.0}{\includegraphics[angle=270,width=0.5\textwidth]{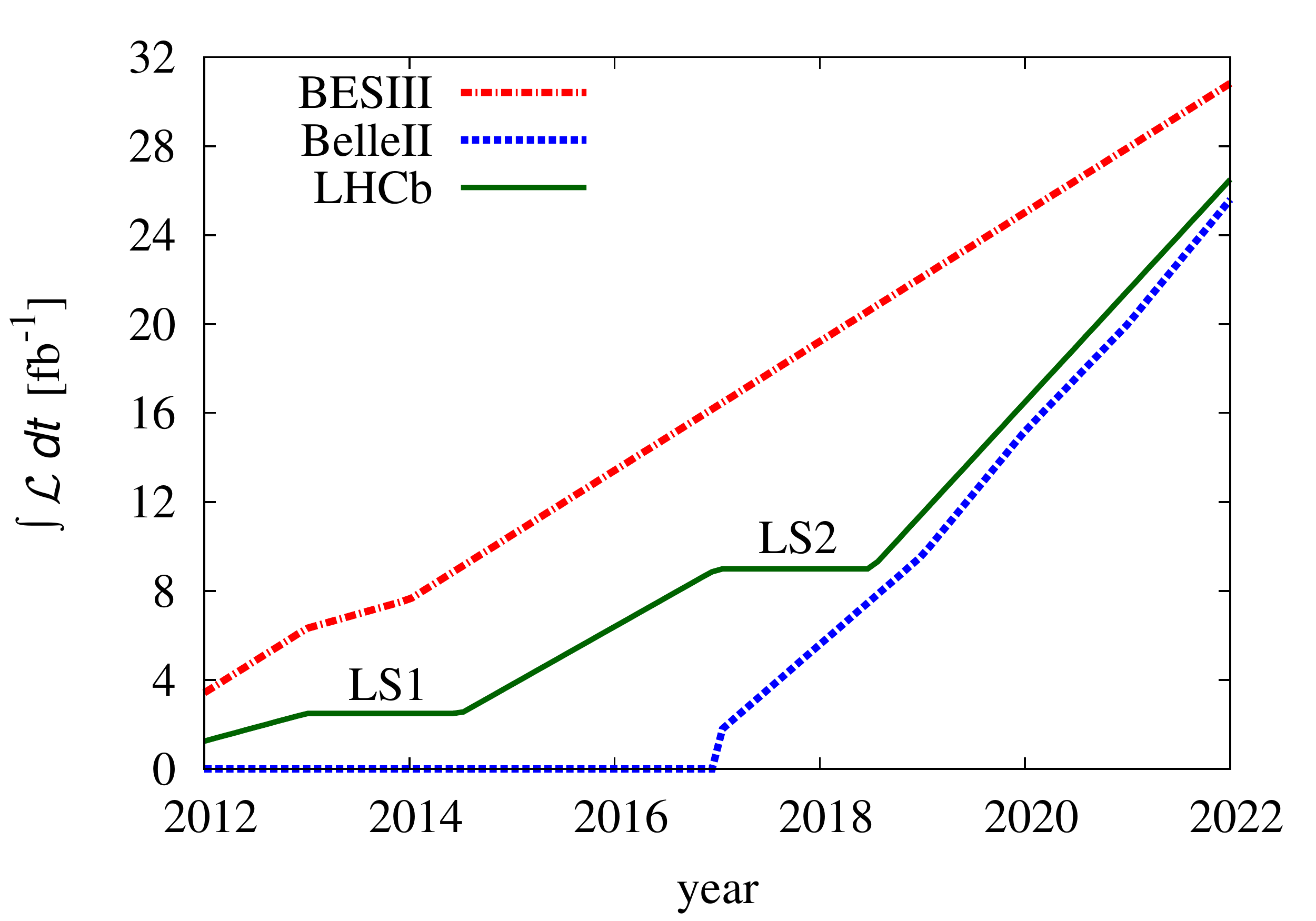}}}
{\scalebox{1.0}{\includegraphics[angle=270,width=0.5\textwidth]{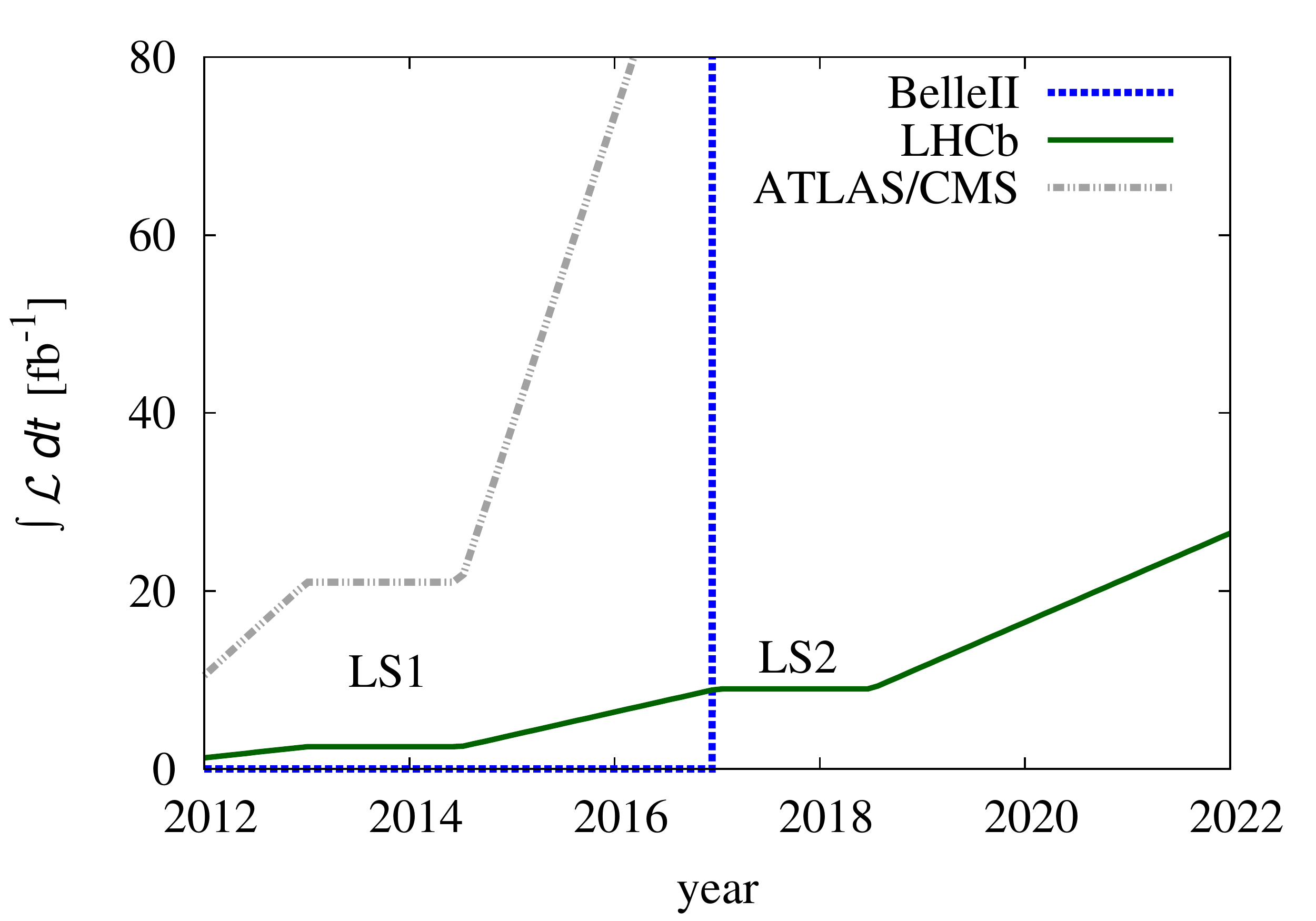}}}
\caption{ Integrated luminosities based on projected run plans for (left) charm and (right) bottom flavor physics experiments~\cite{BESIIIplan,BelleIIplan,LHCbplan,ATLASplan,CMSNote}.  At the time of this review, we are approaching the end of the first long shutdown (LS1) at the LHC.  The BelleII integrated luminosity in the left panel is normalized to LHCb by the expected number of reconstructed benchmark decays, $D^{+*}\to D^0\pi^+$ and $D^0\to K^-\pi^+$.}
\label{fig-intL}
\end{figure}

 A considerable collection of heavy-flavor experimental results have been amassed by 
 the flavor factories (BaBar, Belle, CLEO, and BES), experiments at LEP and the LHC at CERN, a host of Fermilab experiments, and earlier efforts at DESY and SLAC.  
The Heavy Flavor Averaging Group (HFAG) has compiled, and averaged where appropriate, these numerous experimental results~\cite{HFAG}.
With plans to eclipse existing data sets, several flavor experiments are currently underway or are scheduled to begin in the near future.
Projected integrated luminosities for this next wave of charm and bottom experiments are shown in Fig.~\ref{fig-intL}.
These plots foretell a data rich era for heavy flavor physics.  
To fully leverage these data, lattice results with comparable precision are needed.

Having multiple experimental results allows us to extract a single CKM matrix element in multiple ways to search for inconsistencies.
It also allows us to combine experimental data and theory input in a global fit to test the unitarity of the CKM matrix.
These tests provide nontrivial consistency checks of the CKM paradigm of the SM.  
The hope, of course, is to find a process, or a class of processes, in tension with the rest\,---\,pointing us in the direction of potential new physics.
Despite extensive effort~\cite{HFAG, FLAG, UTfit, CKMfitter} there are precious few tensions.
The overall success of the CKM paradigm, combined with the lack of obvious non SM physics in high energy searches, has generated increased interest in rare processes.

\section{CKM Physics}
\label{sec-CKM}
Processes that occur at tree-level in the SM allow a clean determination of CKM matrix elements.
Ideally, we determine a CKM matrix element from multiple processes (e.g. $|V_{ub}|$ from $B\to X_u l \nu$, $B\to\pi l \nu$, $B\to\tau\nu$, $B_s\to Kl\nu$, $\Lambda_b\to pl\nu$, ...) and check for consistency.
The study of these processes also allows us to verify unitarity of the CKM matrix by overdetermining the sides and angles of the unitarity triangle(s).
These tests require the combination of results from theory and experiment.
In order to take full advantage of a particularly well-measured quantity, lattice must calculate with comparable precision\,---\,and vice versa.
This review compares the precision of lattice and experiment using decay constants and form factors.
Experimental bands in plots of decay constants and form factors (Figs.~\ref{fig-fD}, \ref{fig-fB}, \ref{fig-f+DKpi}, and~\ref{fig-G1F1}) indicate only precision and are plotted using the central value of the corresponding lattice average.
The search for tension between the SM and experiment is made at the level of extracted CKM matrix elements (Figs.~\ref{fig-VcdVcs}, \ref{fig-BPi_Vub}, and \ref{fig-Vcb_R}).

\subsection{Leptonic Decays}
\label{sec-leptonic}
Leptonic decays of a meson $\mathcal{P}$, with quark content $q_iq_j$, are described by Eq.~(\ref{eq-leptonic}).
This expression relates measured quantities (on the left hand side), what we calculate on the lattice (the decay constant $f_{\mathcal P}$), and the quantity we are after (the CKM matrix element $|V_{ij}|$).  
Higher order electroweak corrections in this, and all decays we will discuss, are either negligible or are perturbatively accounted for.
\begin{equation}
\mathcal{B}(\mathcal{P}\to  l\nu) \left[ \frac{G_F^2}{8\pi}\ \tau_{\mathcal P}M_{\mathcal P}\ m_ l^2 \Big(1-\frac{m_ l^2}{M^2_{\mathcal P}}\Big)^2 \right]^{-1} =\  f^2_{\mathcal P}\ |V_{ij}|^2 \ +\ \mathcal{O}\Big(\frac{M_{\mathcal P}}{M_W}\Big)^2
\label{eq-leptonic}
\end{equation}

\subsubsection{$D_{(s)}$ leptonic decays}
\label{sec-Dleptonic}
\begin{figure}[t!]
{\scalebox{0.87}{\includegraphics[angle=0,width=0.5\textwidth]{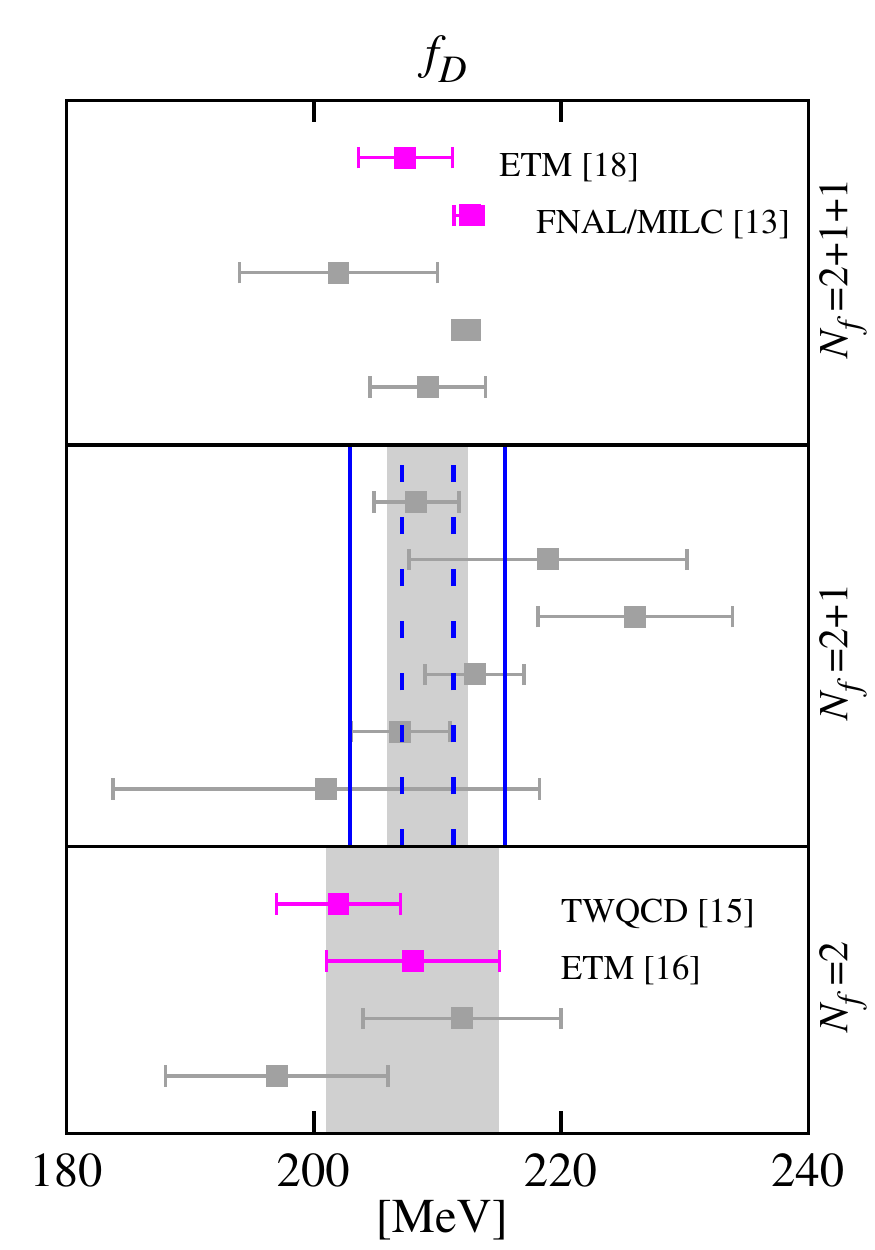}}}
\hspace{0.6in}
{\scalebox{0.87}{\includegraphics[angle=0,width=0.5\textwidth]{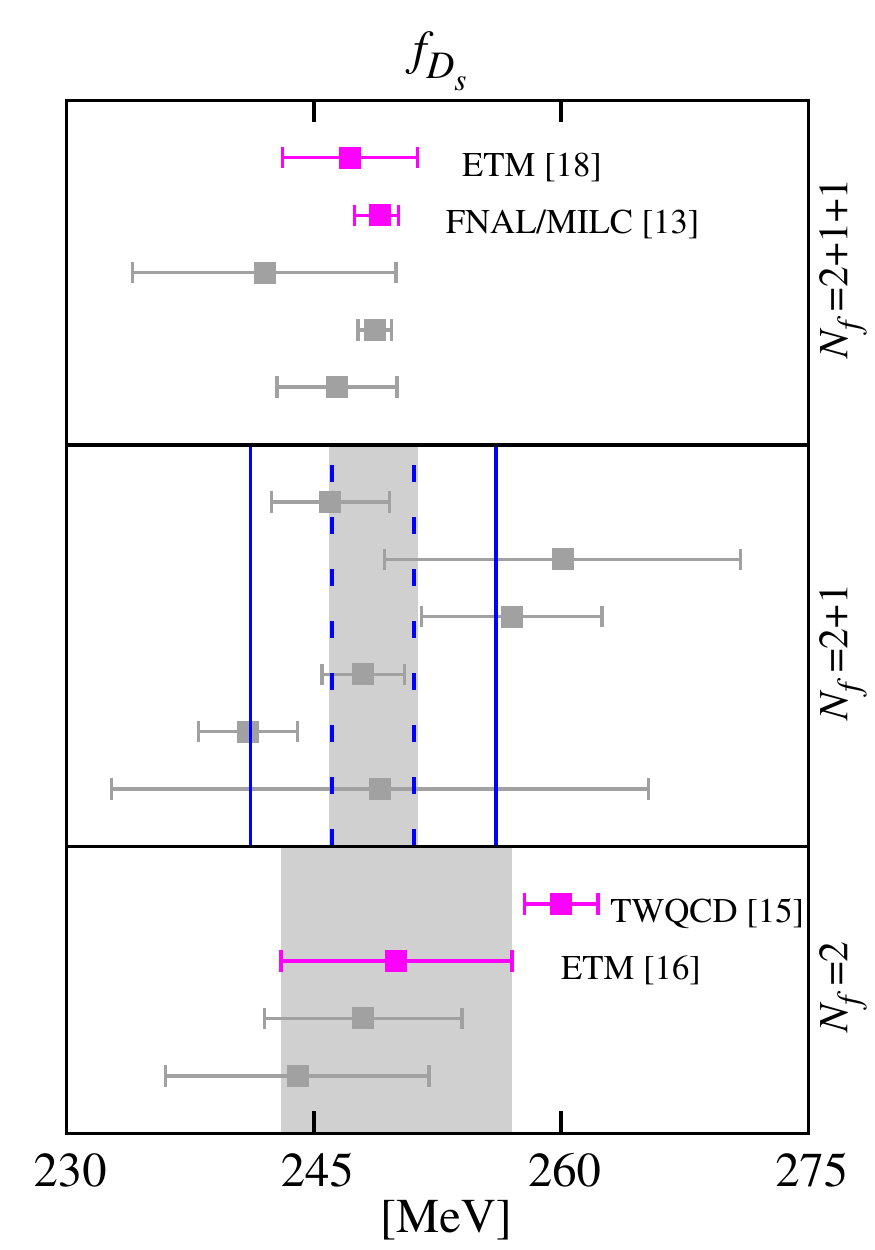}}}
\caption{Recent lattice results are shown in magenta (FNAL/MILC~\cite{Javad:2014}, TWQCD~\cite{TWQCD:2014}, and ETM~\cite{ETM:2014, ETM:2014a}).  Previous year's results, and their averages, are shown in gray and are taken from FLAG~\cite{FLAG}.  The width of the solid blue band indicates current equivalent experimental precision~\cite{HFAG, BESIII-Dtomunu, CLEO} while the narrower dashed blue band shows the projected precision in the year 2020~\cite{BelleII-future, BESIII-future}.}
\label{fig-fD}
\end{figure}

Including the new BESIII result for $\mathcal{B}(D\to\mu\nu)$~\cite{BESIII-Dtomunu}, the branching fractions of $D_{(s)}$ meson decays are known to a precision of $5-5.5\%$.  
Lifetimes of the $D_{(s)}$ mesons are known to better than $1.5\%$ and remaining quantities on the left hand side to better than $0.4\%$ --- branching fractions are the leading source of experimental uncertainty.
In 2020, BelleII~\cite{BelleII-future} and BESIII~\cite{BESIII-future} project branching fraction measurements at the level of $2\%$ and lifetimes to a precision of $1\%$.
Translating these experimental precisions to target errors on the decay constants gives $3\%$ today and $1\%$ in 2020.
Fig.~\ref{fig-fD} plots decay constant values calculated from the lattice community together with bands indicating the corresponding experimental precision.  

Recent lattice works include a calculation by FNAL/MILC~\cite{Javad:2014} using the MILC $N_f=2+1+1$ HISQ configurations with HISQ valence physical-mass light and charm quarks at 0.06, 0.09, 0.12, and 0.15~fm.
They are also calculating~\cite{Ethan} both $D$ and $B$ decay constants using MILC $N_f=2+1$ asqtad configurations with asqtad light quarks and Fermilab charm/bottom quarks.  This calculation includes pions as light as 190~MeV and is being done at 0.045, 0.06, 0.09, 0.12, and 0.15~fm.
TWQCD recently reported results~\cite{TWQCD:2014} using two flavors of domain wall sea quarks with domain wall valence quarks at 0.06~fm and pion masses as light as 260~MeV.
ETM recently calculated~\cite{ETM:2014} $D$ and $B$ decay constants and $B^0_{(s)}$ meson mixing parameters using two flavors of twisted mass sea quarks with an automatically $\mathcal{O}(a)$ improved Symanzik gauge action.
The calculation used twisted mass valence light quarks with pions as light as 280~MeV, four lattice spacings ranging from $0.05-0.10$~fm, and the ratio method to iterate from charm to bottom.
They are extending this work to include physical-mass light quarks~\cite{Bartosz}.
Since the lattice conference they have also reported results~\cite{ETM:2014a} using $N_f=2+1+1$ Iwasaki gauge fields with Wilson twisted mass sea quarks,
three lattice spacings (0.06, 0.08, and 0.09 fm),
pions as light as 210 MeV,
and using Wilson twisted mass light and Osterwalder-Seiler strange and charm valence quarks.
The Southampton-Edinburgh-KEK Collaboration reported~\cite{Tobias} on a pilot study of domain wall charm discretization effects in quenched calculations at 0.035, 0.05, 0.07, and 0.10~fm.  
Based on the results of this pilot study, RBC/UKQCD is using domain wall charm valence quarks on $N_f=2+1$ domain wall configurations at 0.115, 0.085, and 0.065~fm with physical-mass light quarks.
They plan to calculate a host of charm physics observables and have plans to extrapolate to bottom quark masses~\cite{Juettner}.

\subsubsection{$B_{(s)}$ leptonic decays}
\label{sec-Bleptonic}
\begin{figure}[t!]
{\scalebox{0.87}{\includegraphics[angle=0,width=0.5\textwidth]{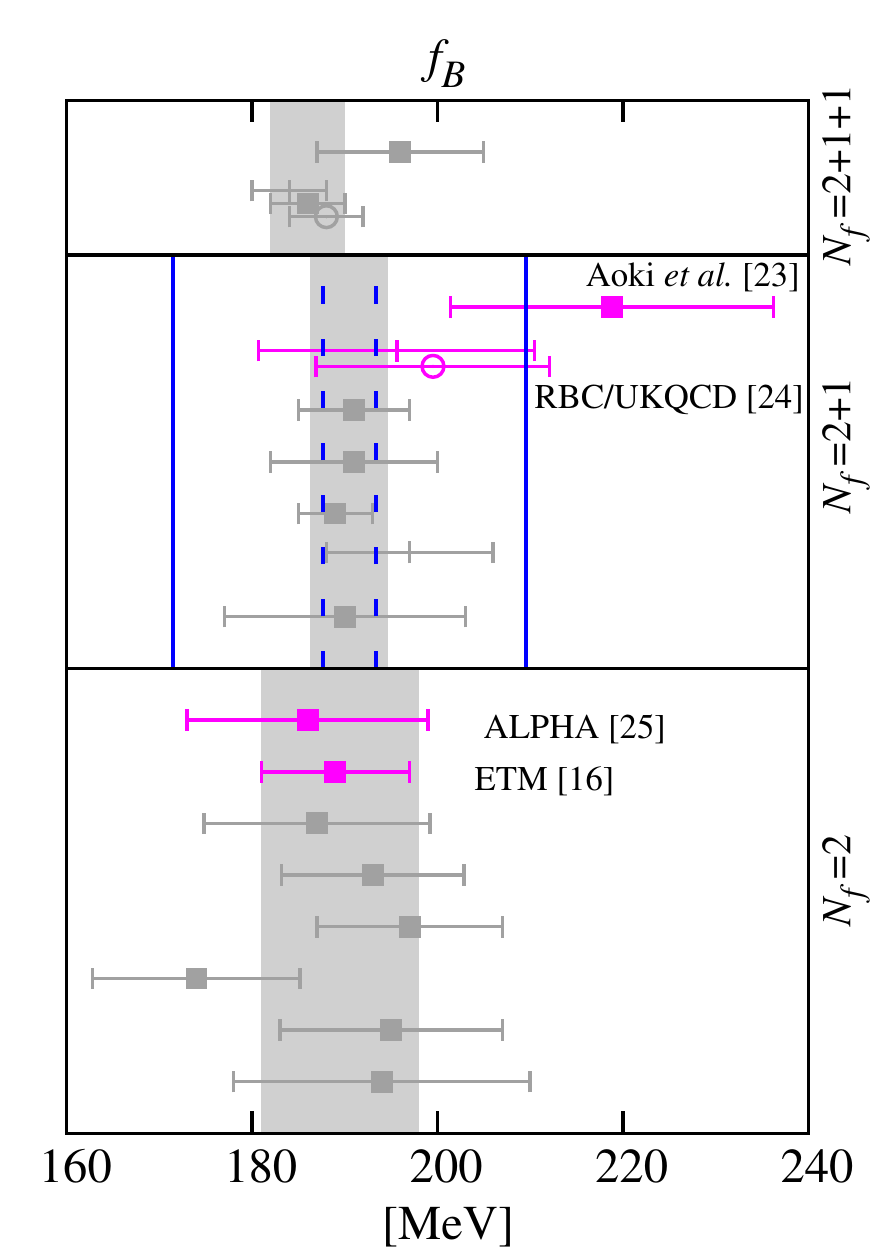}}}
\hspace{0.6in}
{\scalebox{0.87}{\includegraphics[angle=0,width=0.5\textwidth]{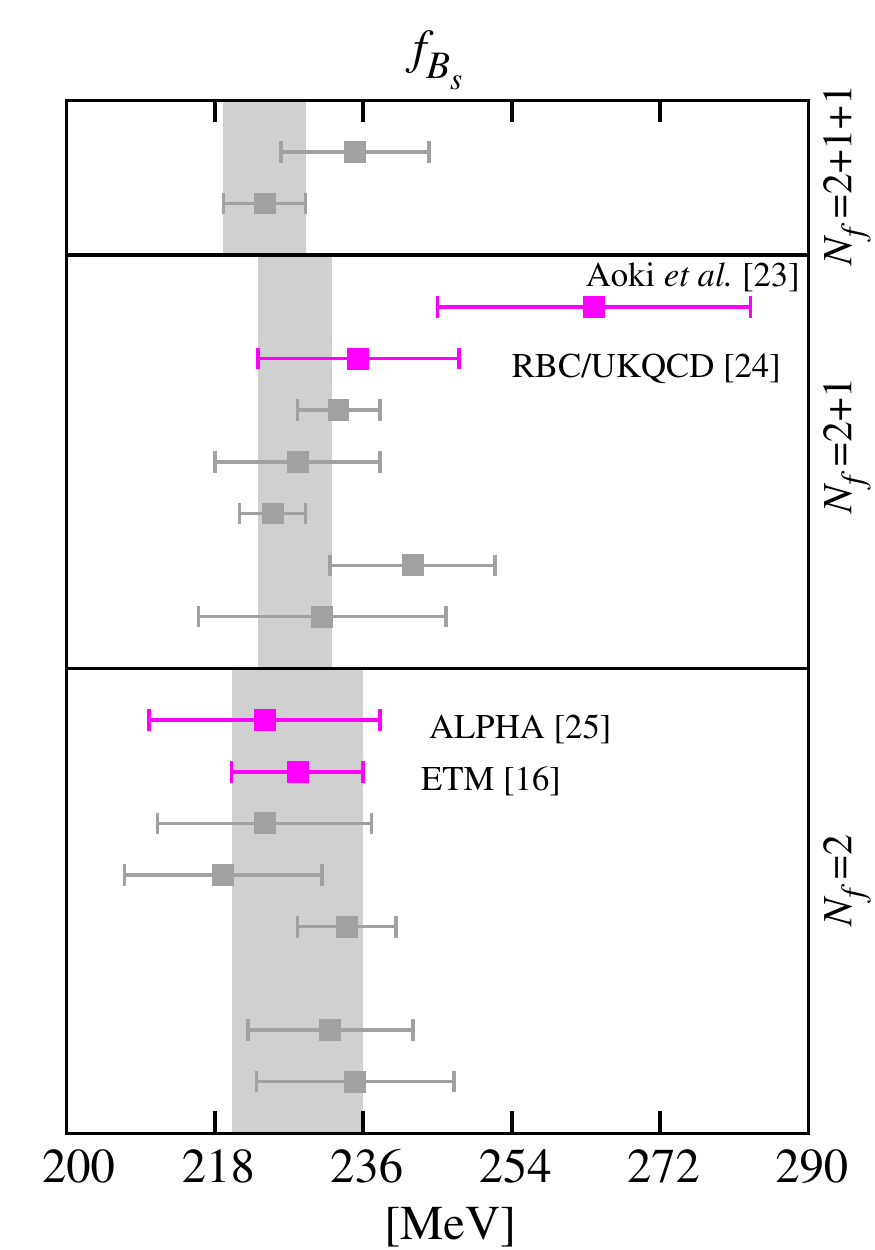}}}
\caption{ Recent lattice results are shown in magenta (Aoki {\it et al.}~\cite{Aoki}, RBC/UKQCD~\cite{RBC/UKQCD}, ALPHA~\cite{ALPHA}, and ETM~\cite{ETM:2014}).  Previous year's results and their averages are shown in gray and are taken from FLAG~\cite{FLAG}.  RBC/UKQCD report separate results for chiral extrapolations to the $B^0$ (circle) and $B^\pm$ (plus).  The solid blue band indicates current equivalent experimental precision~\cite{HFAG} while the narrower dashed blue band gives the projected precision in the year 2020~\cite{BelleII-future}.}
\label{fig-fB}
\end{figure}
The branching fraction for the leptonic decay $B\to\tau\nu$ has been measured to $20-25\%$ precision~\cite{HFAG} and $B$ meson lifetimes are known to better than $0.5\%$.  As with $D$ leptonic decays, the branching fraction dominates the experimental uncertainty.  BelleII projects~\cite{BelleII-future} a significantly improved $3\%$ measurement of the branching fraction by 2020.  The corresponding target uncertainties on $f_B$ are $10\%$ now and $1.5\%$ in 2020.
Fig.~\ref{fig-fB} plots recent lattice calculations of $f_{B_{(s)}}$ and current and projected target uncertainties for $f_B$.
The fictitious $B_s$ decay constant is useful in ratios and in the parameterization of hadronic inputs, e.g. in the SM prediction of $B_s\to\mu^+\mu^-$~\cite{BurasfBs}.
Despite the favorable comparison of lattice results with experiment, the role of $B$ decay constants in parameterizing hadronic uncertainties in other processes makes their improvement a continuing priority.

Recent lattice works include a calculation by Aoki {\it et al.}~\cite{Aoki} in the static limit.
These results will anchor a heavy quark expansion, guided by results using the relativistic heavy quark action near the charm mass, and iterating to the anchor point via the ratio method.  They simulate with $N_f=2+1$ domain wall sea quarks and Iwasaki gauge fields, pion masses as light as 289~MeV, one-loop matching including $\mathcal{O}(a)$ effects, and at lattice spacings of 0.09 and 0.11~fm.  
Planned improvements include working with physical-mass light quarks and performing non-perturbative renormalization via RI-MOM.  Their results are plotted in Fig.~\ref{fig-fB} with and without estimated $\mathcal{O}(1/m_b)$ errors.
RBC/UKQCD calculated $f_{B_{(s)}}$ using $N_f=2+1$ domain wall sea quarks on Iwasaki gauge fields with domain wall valence light quarks and a nonperturbatively-tuned relativistic heavy quark action for the $b$ quark~\cite{RBC/UKQCD}.  
Their simulations include pion masses down to 290~MeV and lattice spacings of 0.09 and 0.11~fm.
ALPHA recently reported~\cite{ALPHA} results of a calculation with $\mathcal{O}(a)$ improved Wilson light valence and $N_f=2$ sea quarks, Wilson gauge action, and HQET treatment of the $b$ quark with non-perturbative NLO improvement.  They simulate at pion masses as light as 190~MeV and at lattice spacings of 0.05, 0.065, and 0.075~fm.
The ETM calculation discussed above for $D$ meson decay constants~\cite{ETM:2014} also included results for $f_{B_{(s)}}$.

\subsection{Semileptonic Decays}
\label{sec-semileptonic}
The semileptonic decay of parent meson $\mathcal{P}$ into daughter meson $\mathcal{D}$ via the quark flavor-changing interaction $q_i\to q_f$ is described by
\begin{equation}
\frac{d\mathcal{B}(\mathcal{P}\to \mathcal{D} l\nu) }{dq^2} \frac{24 \pi^3}{\tau_{\mathcal P} G_F^2 |{\bf p}_{\mathcal D} |^3}  = |V_{if}|^2 \  |f_+|^2 + \mathcal{O}\Big( \frac{m_l^2}{q^2} \Big).
\label{eq-SL}
\end{equation}
Unlike leptonic decays, semileptonic decays have a kinematic dependence associated with the four-momentum $q^2$ carried away by the lepton-neutrino pair. 
Contributions from the scalar form factor $f_0$ are suppressed by $m_l^2/q^2$ and are negligible provided experimental data used in the combination are appropriately restricted in $q^2$.
Lattice simulations are typically performed in the parent meson rest frame with low daughter meson momenta, i.e. large $q^2$.
To determine form factors over the full kinematic range, $m_l^2 \leq q^2 \leq (M_{\mathcal{P}}-M_{\mathcal{D}})^2$, lattice results must be extrapolated to low $q^2$.
The extrapolation is significant, and is therefore more of an issue, for semilieptonic $B_{(s)}$ decays.
The $z$ expansion~\cite{zexp} allows the extrapolation to be carried out in a model-independent way.
Ref.~\cite{Bouchard:BsK} combines the $z$~expansion with hard pion chiral perturbation theory~\cite{hpchpt} in an approach that may permit direct lattice calculation of form factors at low $q^2$, a possibility being explored in~\cite{Bouchard:BPi}.
Branching fraction measurements in $q^2$ bins combined with the calculated $q^2$-dependence of form factors allows shape comparison and improves the extraction of $|V_{if}|$.

\subsubsection{$D_{(s)}$ semileptonic decays}
\label{sec-Dsemileptonic}
\begin{figure}[t!]
{\scalebox{0.87}{\includegraphics[angle=0,width=0.5\textwidth]{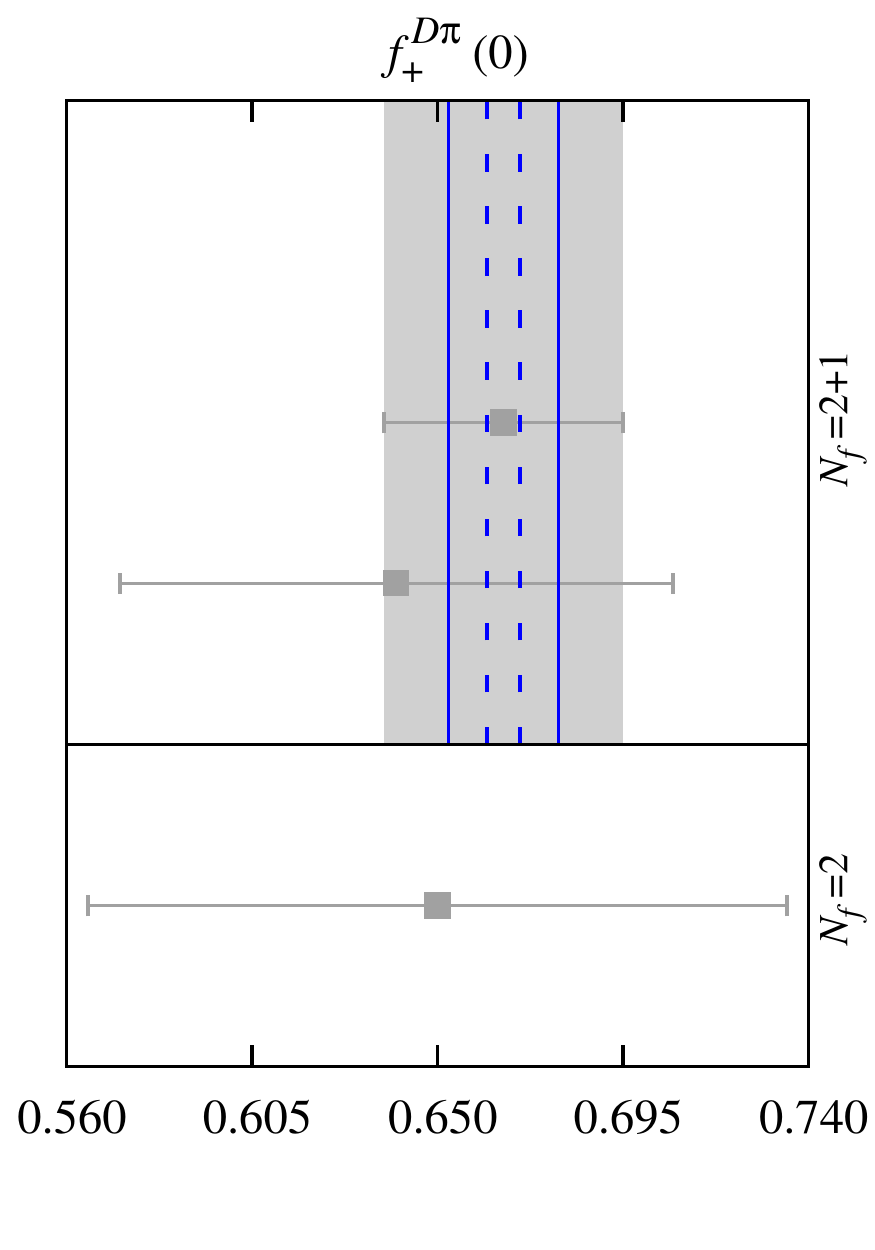}}}
\hspace{0.6in}
{\scalebox{0.87}{\includegraphics[angle=0,width=0.5\textwidth]{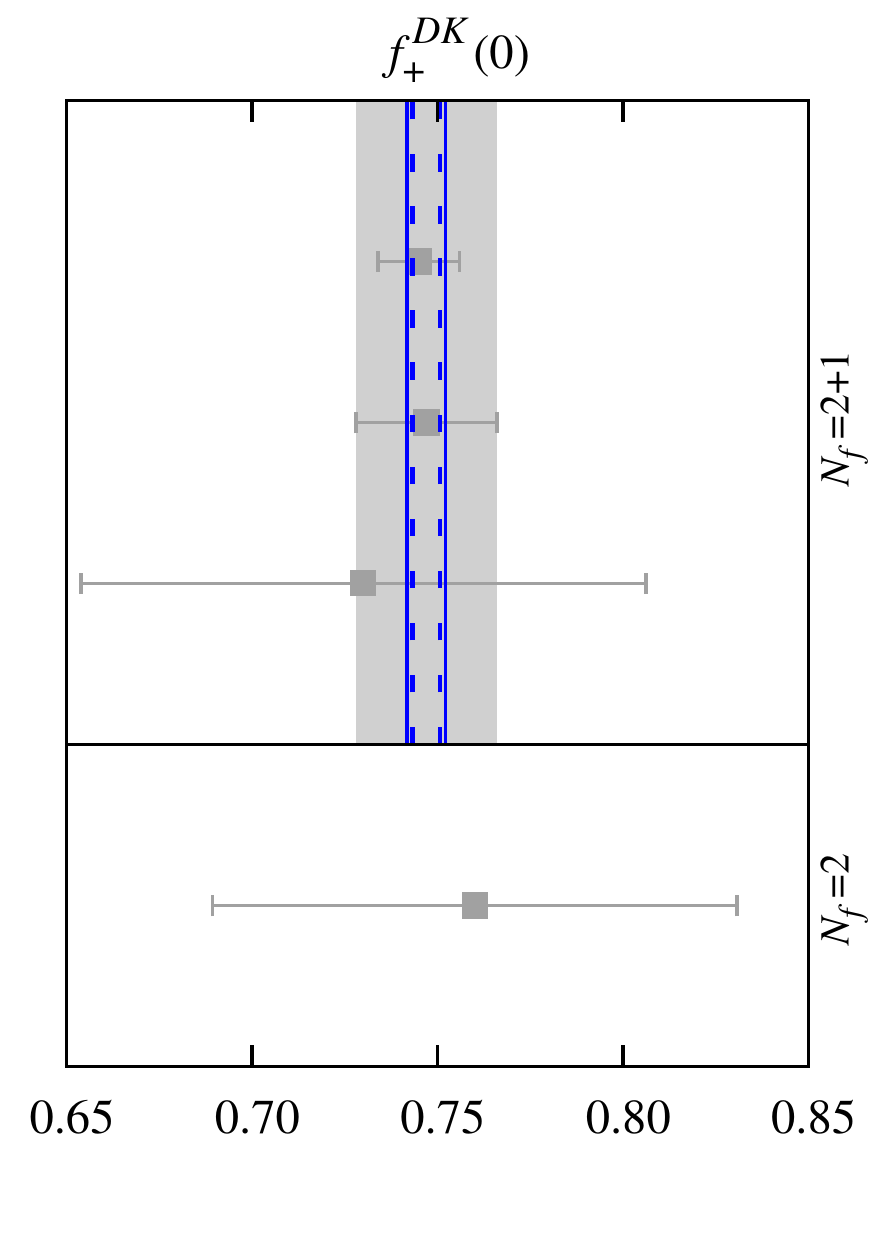}}}
\vspace{-0.25in}
\caption{ Previous year's results and their averages are shown in gray and are taken from Refs.~\cite{FLAG, Koponen:2013}.  Solid blue bands indicate current equivalent experimental precision~\cite{HFAG} while the narrower dashed blue bands give the projected precision in the year 2020~\cite{BelleII-future, BESIII-future}.}
\label{fig-f+DKpi}
\end{figure}

The semileptonic decays $D\to\pi l\nu$ and $D\to Kl\nu$ ($l=e, \mu$; decay to a tau is not energetically allowed) permit the determination of the CKM matrix elements $|V_{cd}|$ and $|V_{cs}|$.
Current and projected experimental precisions for $\mathcal{B}(D\to\pi e\nu)$ and $\mathcal{B}(D\to Ke\nu)$ give targets for form factor calculations shown in Fig.~\ref{fig-f+DKpi}.%
\footnote{$D$ semileptonic determinations of $|V_{cd}|$ and $|V_{cs}|$ have been largely limited to extrapolated values at $q^2=0$.  
Experimental results are typically recast as form factors using models for the $q^2$ dependence.  
The lattice efforts underway plan to calculate the form factors at $q^2>0$.
It would behoove us to communicate to our experimental colleagues that future measurements would best be reported in a model independent way, i.e. as branching fractions or decay rates in $q^2$ bins.}
Relative to experiment, lattice has plenty of room for improvement in $D$ semileptonic decays.
Though there were no results in the past year, there are several efforts underway.
A FNAL/MILC calculation is utilizing the MILC $N_f=2+1$ asqtad ensembles,
asqtad light and strange valence quarks,
a FNAL charm quark, and
four lattice spacings (0.045, 0.06, 0.09, and 0.12 fm).
They plan to form ratios of form factors with $D_{(s)}$ decay constants to cancel charm quark discretization effects, then combine the ratios with the high-precision results of Ref.~\cite{Javad:2014}.
A separate effort by FNAL/MILC~\cite{Thom} uses the MILC $N_f=2+1+1$ HISQ ensembles,
HISQ valence quarks,
three lattice spacings (0.06, 0.09, and 0.12 fm), and
physical quark masses.
ETM is calculating the form factors using $N_f=2+1+1$ twisted mass gauge fields, three lattice spacings, and pion masses as light as 210 MeV~\cite{Lorenzo}.

The $D_s\to\phi l\nu$ decay provides an alternative to $D\to Kl\nu$ in the determination of $|V_{cs}|$.  
The first unquenched calculation of the axial and vector form factors for this decay, and their combination with experiment to extract $|V_{cs}|$, was performed by HPQCD in Ref.~\cite{Donald}.
Given expected improvements at BESIII, theory errors dominate this determination of $|V_{cs}|$.
Kanamori {\it et al.} are performing an exploratory calculation of $D_s\to\eta^{(} {'} {}^{)}l\nu$ using the QCDSF $N_f=2+1$ stout link ensembles, 
two $SU(3)_{\rm flavor}$ symmetric sets of quark masses ($M_\pi = M_\eta = 370\ {\rm and}\ 470\ {\rm MeV})$,
and $a=0.075\ {\rm fm}$.
They are calculating $\eta-\eta '$ mixing angles and including disconnected contributions to $D_s\to\eta' l\nu$~\cite{Kanamori}.

\begin{figure}[t!]
{\scalebox{0.9}{\includegraphics[angle=0,width=0.5\textwidth]{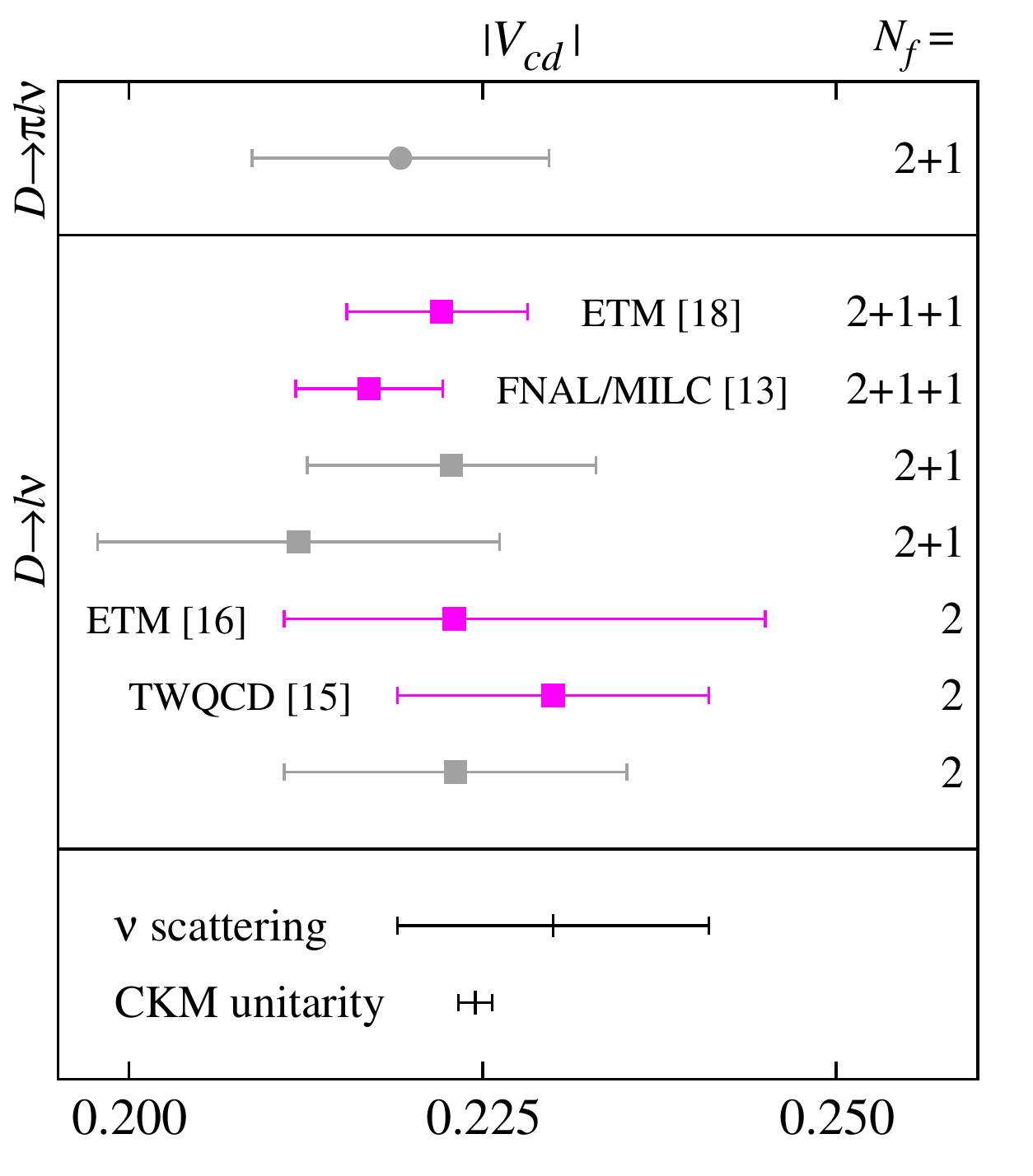}}}
\hspace{0.5in}
{\scalebox{0.9}{\includegraphics[angle=0,width=0.5\textwidth]{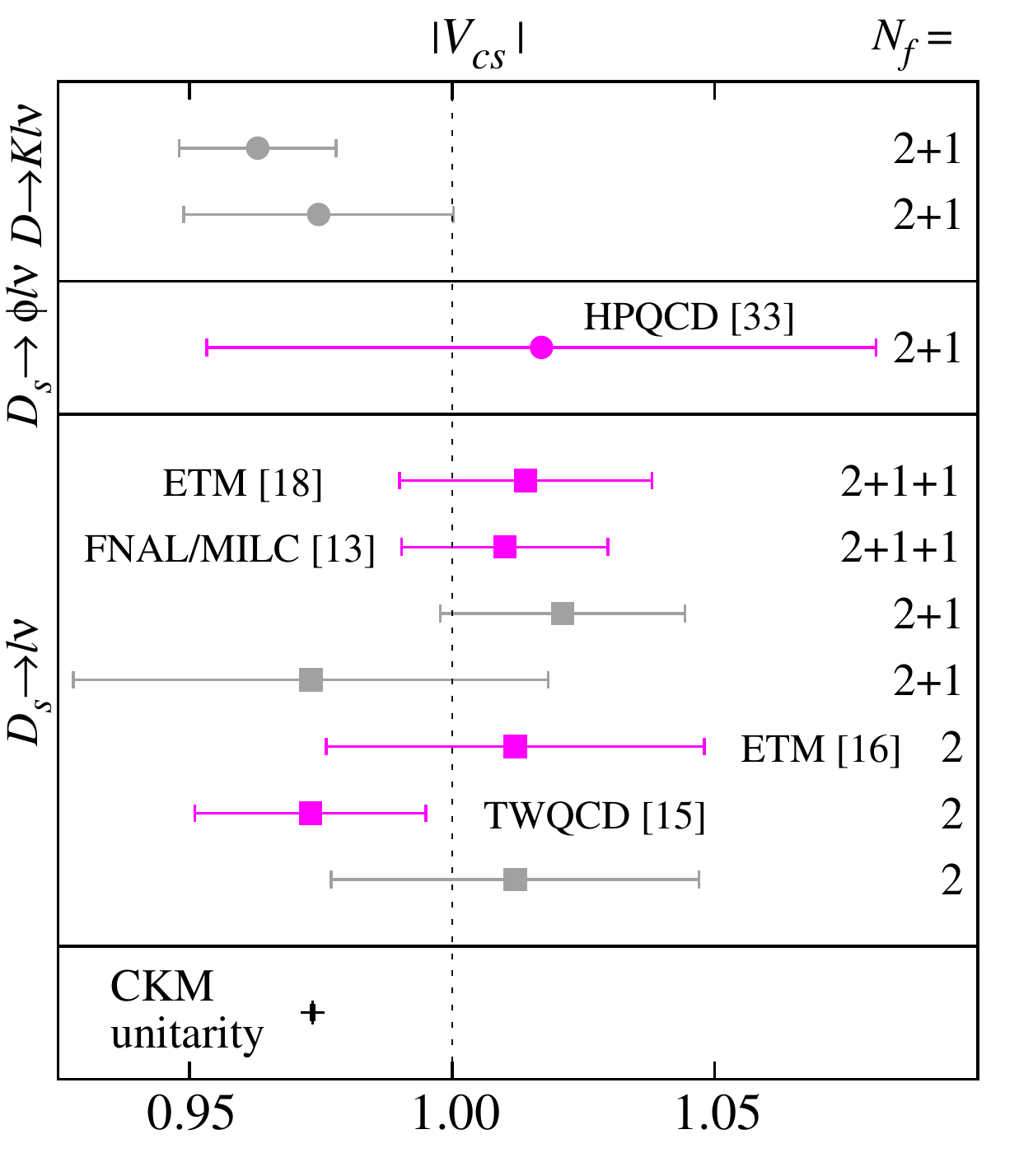}}}
\caption{ Recent lattice results from FNAL/MILC~\cite{Javad:2014}, ETM~\cite{ETM:2014, ETM:2014a}, TWQCD~\cite{TWQCD:2014}, and HPQCD~\cite{Donald} are shown in magenta.  Previous year's results are shown in gray and are taken from FLAG~\cite{FLAG}.}
\label{fig-VcdVcs}
\end{figure}

A collection of leptonic, semileptonic, and other determinations of $|V_{cd}|$ and $|V_{cs}|$ are plotted in Fig.~\ref{fig-VcdVcs}.
At the current level of precision the different determinations of $|V_{cd}|$ are in very good agreement and are all in agreement with the assumption of CKM unitarity.
Given current and projected levels of experimental precision in $D\to\pi l\nu$, improvement in the corresponding lattice form factors for this decay is desirable.
There is a slight tension between values of $|V_{cs}|$ obtained from leptonic $D_s$ decays and those from $D\to Kl\nu$ and unitarity.
Given this tension, it would be particularly interesting to improve upon the theory error associated with $D_s\to\phi l\nu$ (see Ref.~\cite{Donald} for discussion) to fully leverage upcoming BESIII results for this decay.

\subsubsection{$B_{(s)}$ semileptonic decays}
\label{sec-Bsemileptonic}
\begin{figure}[t!]
\hspace{0.25in}{\scalebox{1.0}{\includegraphics[angle=0,width=0.5\textwidth,bb=85 372 585 600]{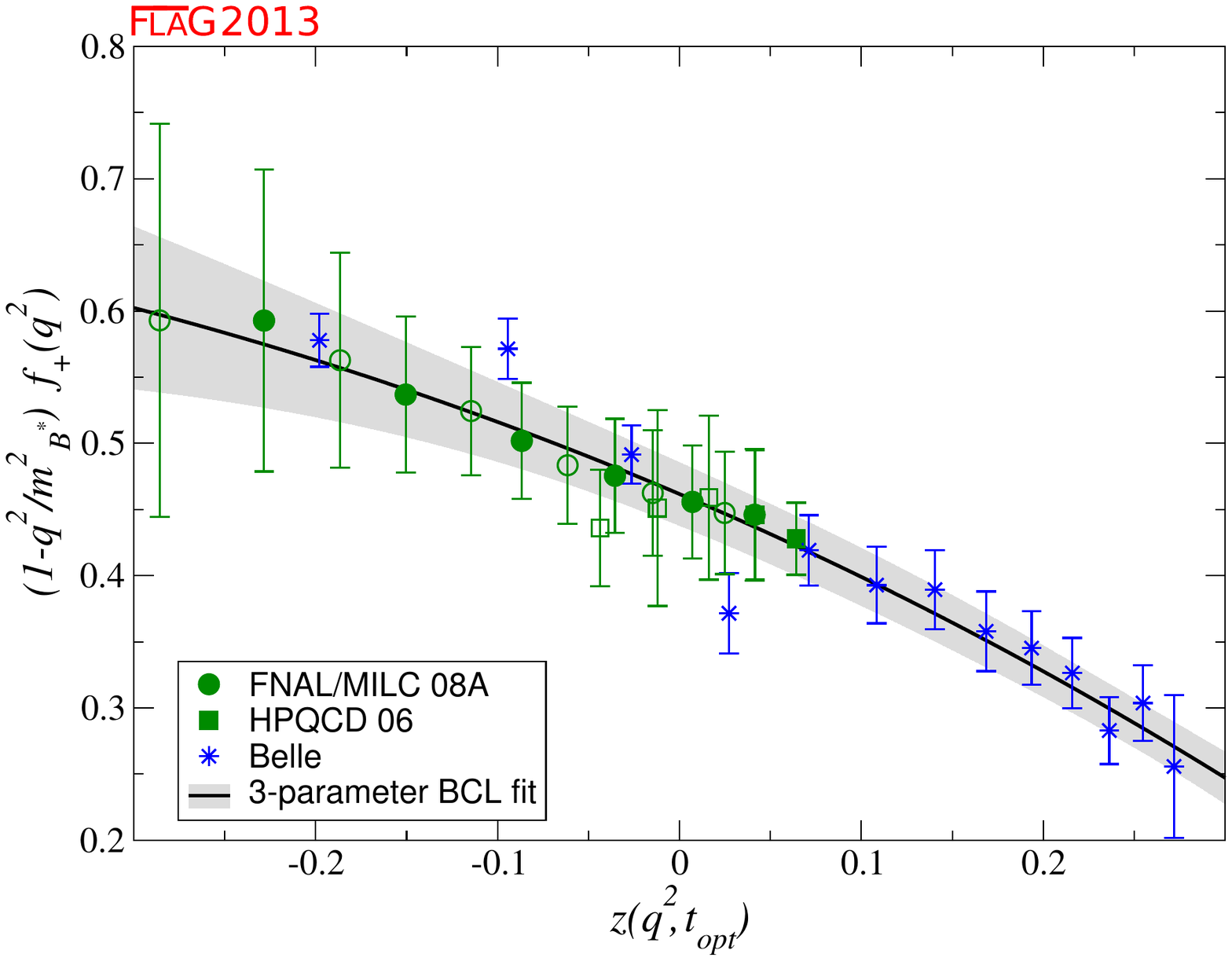}}}
\hspace{0.1in}
{\scalebox{0.9}{\includegraphics[angle=0,width=0.5\textwidth]{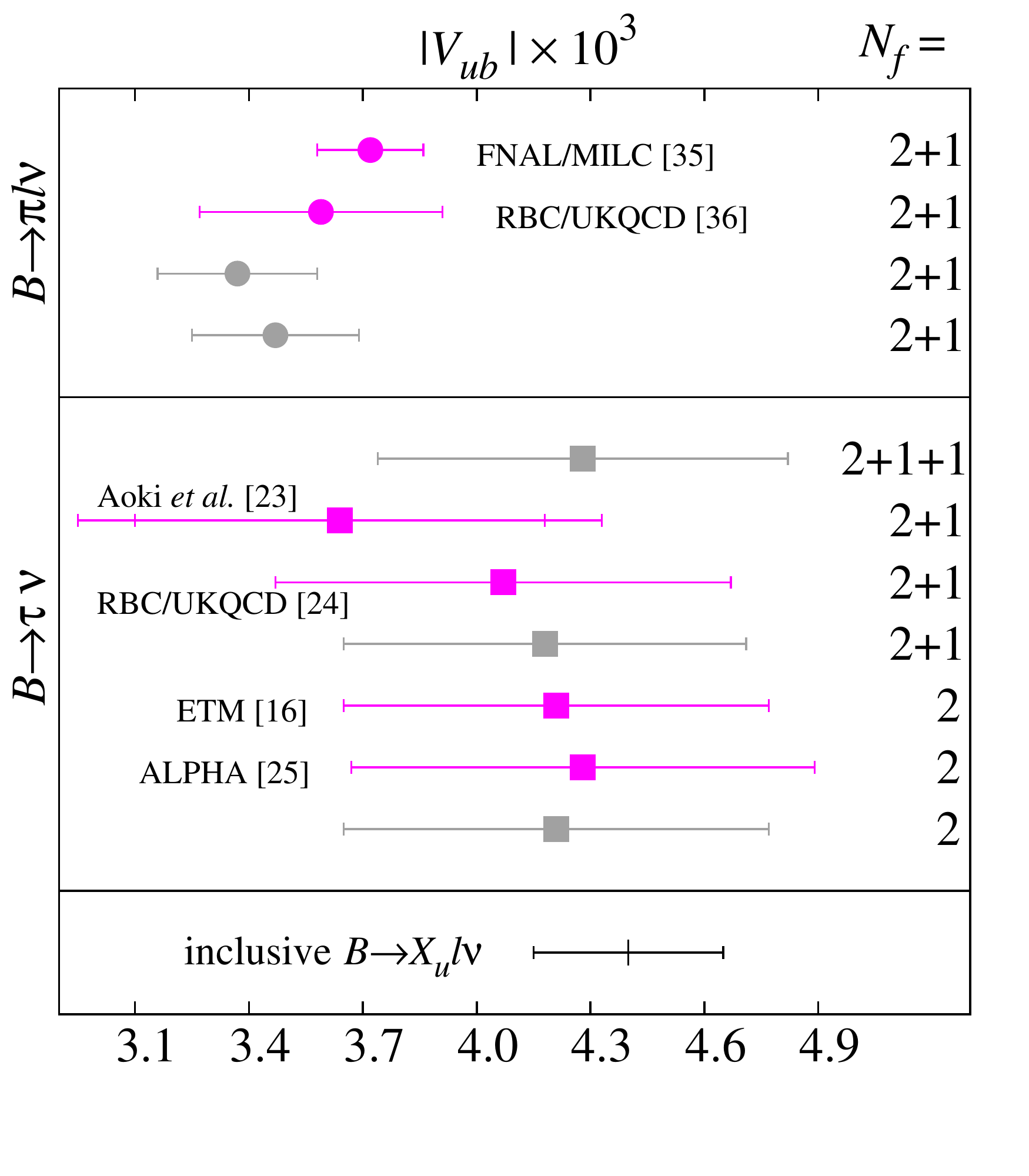}}}
\vspace{-0.4in}
\caption{(left) FLAG's~\cite{FLAG} simultaneous fit of Belle $B\to\pi l\nu$ data (blue) and previous year's lattice data (green) illustrates the current state of precision between lattice and experiment.  Experimental results have errors roughly half the size of lattice errors.  
(right) Recent reported values for $|V_{ub}|$ from FNAL/MILC~\cite{Du} and RBC/UKQCD~\cite{Taichi}, and values of $|V_{ub}|$ obtained from combination of $\mathcal{B}(B\to\tau\nu, {\rm BABAR} + {\rm Belle})$~\cite{FLAG} with recent calculations of $f_B$ by Aoki {\it et al.}~\cite{Aoki}, RBC/UKQCD~\cite{RBC/UKQCD}, ALPHA~\cite{ALPHA}, and ETM~\cite{ETM:2014}, are shown in magenta with previous year's results in gray~\cite{FLAG}.}
\label{fig-BPi_Vub}
\end{figure}

The only exclusive semileptonic decay currently used in the determination of $|V_{ub}|$ is \mbox{$B\to\pi l\nu$}.\footnote{Hopefully this will change with measurements of $\mathcal{B}(B_s\to Kl\nu)$ and $\mathcal{B}(\Lambda_b\to pl\nu)$ at LHCb and/or BelleII.}  
The left-hand side of Eq.~(\ref{eq-SL}) is known from experiment to about $8\%$ while lattice has determined the squared form factor on the right-hand side with an error roughly twice as large.
The current experimental error is expected to be cut nearly in half within the first $5\ {\rm ab}^{-1}$ of data at BelleII~\cite{BelleII-future}, i.e. within the first year or two of data taking.
This is a rough characterization of the status of experiment and lattice as the analysis of $B\to\pi l\nu$ is done over the full kinematic range of $q^2$, as shown in the left panel of Fig.~\ref{fig-BPi_Vub}.  This plot, taken from~\cite{FLAG}, shows the result of a simultaneous $z$ expansion fit to lattice and experiment to extract $|V_{ub}|$.

There are several ongoing lattice calculations of the $B\to\pi l\nu$ form factors.
Preliminary results from FNAL/MILC~\cite{Du} have errors matching current experimental precision and yield the most precise determination of $|V_{ub}|$ to date.
This calculation uses the MILC $N_f=2+1$ asqtad configurations with a FNAL $b$ quark and asqtad light valence quarks,
four lattice spacings (0.045, 0.06, 0.09, and 0.12 fm), and
pions as light as 177 MeV.
Recent results from RBC/UKQCD~\cite{Taichi} use $N_f=2+1$ domain wall sea quarks and Iwasaki gauge fields, domain wall light valence quarks, and a non-perturbatively tuned relativistic heavy quark treatment of the $b$ quark.
They simulate at two lattice spacings (0.09 and 0.11 fm) and with pions as light as 289 MeV.
%
HPQCD is calculating form factors for this decay using the MILC $N_f=2+1$ asqtad gauge fields, NRQCD $b$ and HISQ light valence quarks, two lattice spacings (0.09 and 0.12 fm), and pion masses down to 190 MeV~\cite{Bouchard:BPi}.
This work is also exploring the feasibility of using the combination of hard pion chiral perturbation theory and the $z$ expansion to include lattice simulation data at low values of $q^2$.

The right panel of Fig.~\ref{fig-BPi_Vub} shows inclusive and exclusive semileptonic and leptonic determinations of  $|V_{ub}|$.
Using the FNAL/MILC preliminary result, $|V_{ub}|= 3.72(14) \times 10^{-3}$, a $2.4 \sigma$ persistent tension remains between the inclusive and exclusive semileptonic determinations.
Though current experimental errors prevent $|V_{ub}|$ from $B\to\tau\nu$ from rivaling the precision of semileptonic determinations, BelleII 2020 projections for $\mathcal{B}(B\to\tau\nu)$~\cite{BelleII-future} combined with current lattice precision for $f_B$ suggest a leptonic $|V_{ub}|$ with $2\%$ precision in five years.  

The semileptonic decay $B_s\to K l\nu$, which differs from $B\to\pi l\nu$ only in its strange spectator quark, offers an alternative exclusive determination of $|V_{ub}|$.
Though not yet observed, a measurement is underway at LHCb and there are prospects during an $\Upsilon(5S)$ run at BelleII.  
There are published $B_s\to Kl\nu$ form factor results from HPQCD~\cite{Bouchard:BsK}, 
recent results from RBC/UKQCD~\cite{Taichi}, 
and an ongoing calculation by FNAL/MILC.
Because of the similarity with $B\to\pi l\nu$, each of these calculations follows closely the same collaboration's efforts described above for $B\to\pi l\nu$.
In addition to these works, there is an ongoing effort by ALPHA~\cite{Bahr}.

\begin{figure}[t!]
{\scalebox{0.9}{\includegraphics[angle=0,width=0.5\textwidth]{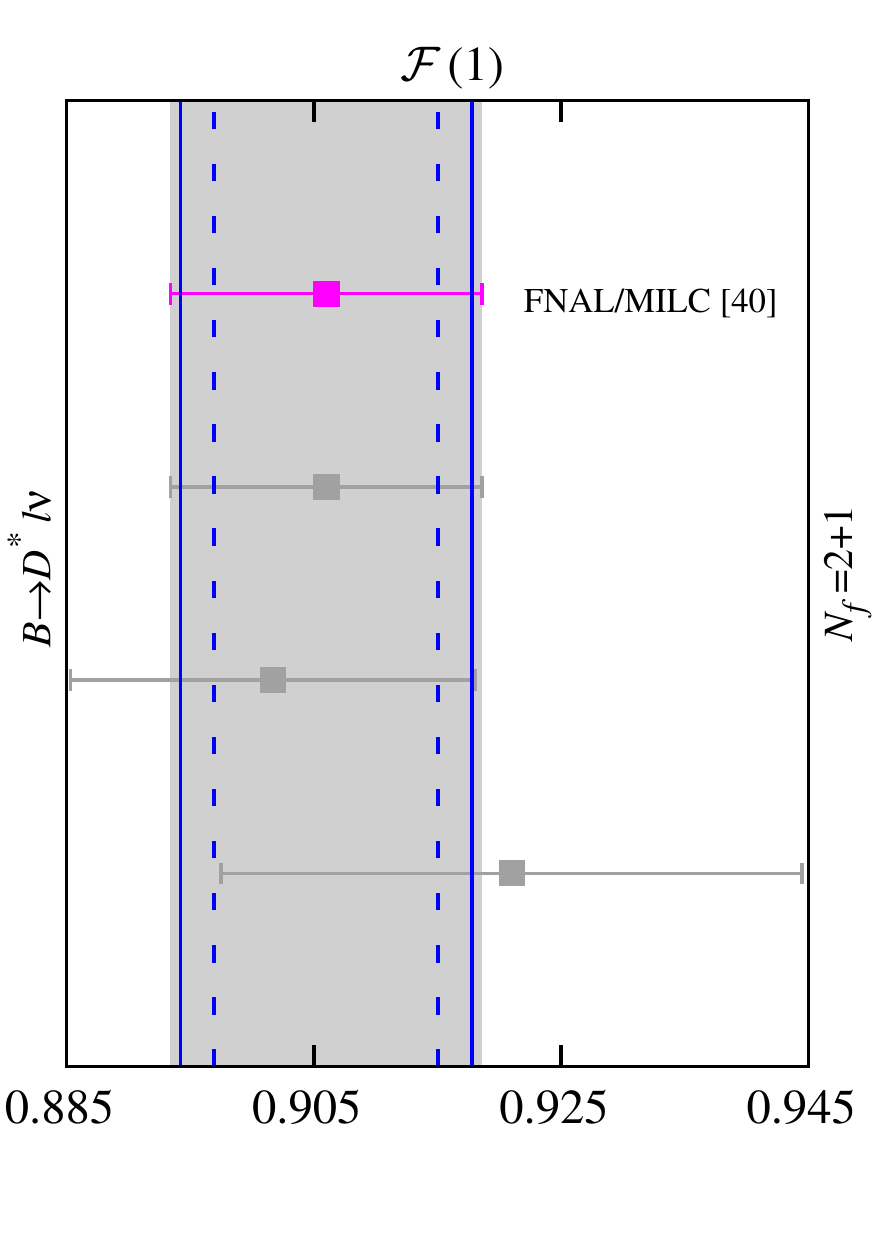}}}
\hspace{0.5in}
{\scalebox{0.9}{\includegraphics[angle=0,width=0.5\textwidth]{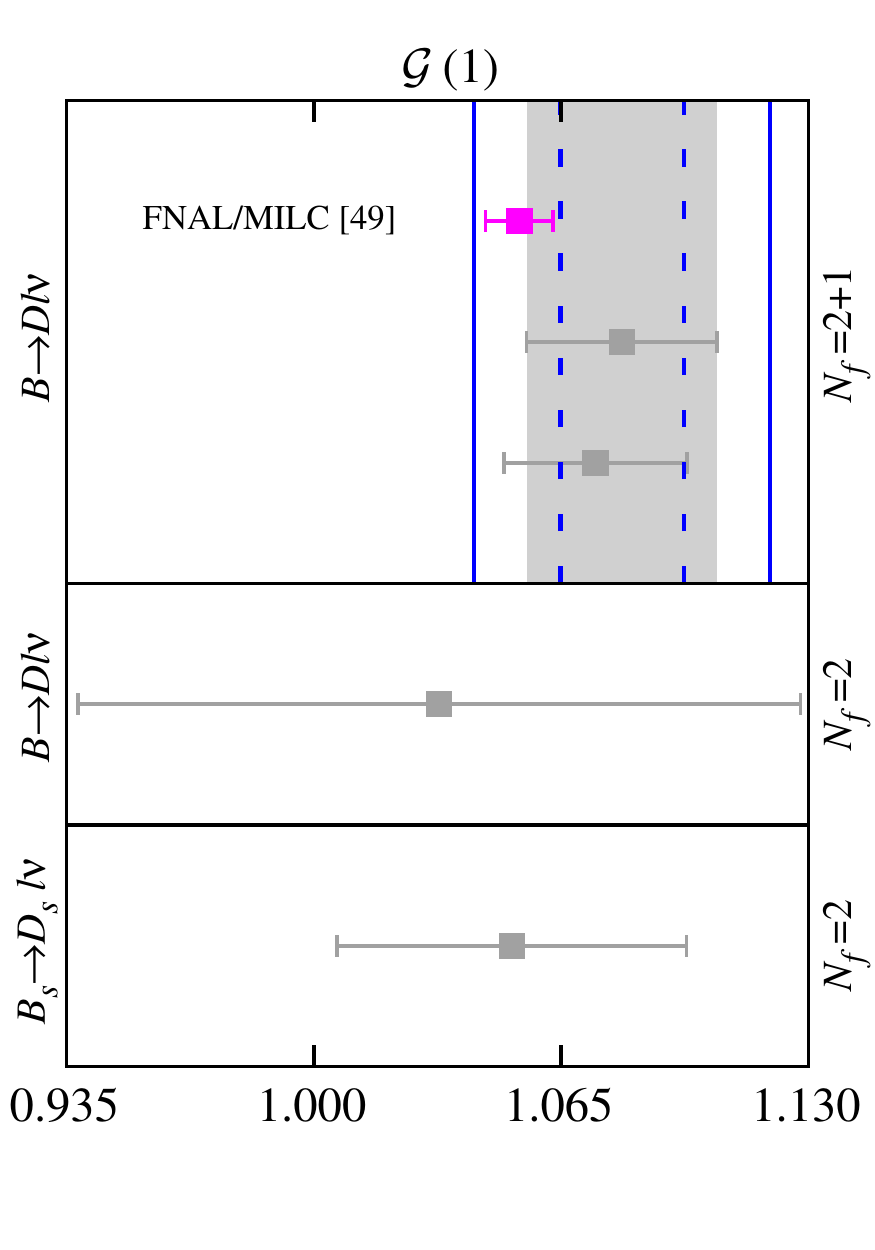}}}
\vspace{-0.5in}
\caption{Comparison of form factors at zero recoil from various lattice calculations.  
Recent results from FNAL/MILC~\cite{FNAL-Vcb, FNAL:BD} are shown in magenta.  
Previous year's results for $B\to D^*l\nu$ and lattice averages (gray bands) and are taken from FLAG~\cite{FLAG}.  
Previous results for $B_{(s)}\to D_{(s)}l\nu$ are taken from Refs.~\cite{Qiu, Okamoto, Atoui}.
The solid blue bands give the current equivalent experimental precision~\cite{HFAG} while the narrower dashed blue bands give the projected equivalent precision in 2020~\cite{BelleII-future}.}
\label{fig-G1F1}
\end{figure}

In addition to the obvious role it plays in tests of CKM unitarity, $|V_{cb}|$ is also important in the determination of several phenomenologically important quantities, e.g. $\mathcal{B}(B_s\to\mu\mu)$~\cite{Bobeth} and $\epsilon_K$~\cite{Buras}.
The primary decay for the exclusive determination of $|V_{cb}|$ is $B\to D^* l\nu$.  For $B^\pm$ (the $B^0$ decay receives an additional correction for final state interactions) the differential decay rate is
\begin{equation}
\frac{d\Gamma (B^\pm\to D^{0*} l\nu)}{dw} \frac{4\pi^3}{G_F^2 |\eta_{EW}|^2 M_{D^*}^3 (M_B-M_{D^*})^2 \sqrt{w^2-1}} =  |V_{cb}|^2 \chi(w) |\mathcal{F}(w)|^2\ +\  \mathcal{O}\left(\frac{m_ l^2}{q^2}\right),
\label{eq-BDstar}
\end{equation}
where the kinematics are described in terms of $w=(p_B/M_B)\cdot(p_{D^*}/M_{D^*})$,
$|\eta_{\rm EW}|$ includes electroweak corrections,
and $\chi(w) |\mathcal{F}(w)|^2$ is the conventional parameterization of the form factors.
Most work is done in the zero recoil ($D^*$ at rest) limit where $w\to1$, $\chi(w)\to1$, and there are additional simplifications (see, e.g., Ref.~\cite{FNAL-Vcb}).
The left-hand side of Eq.~(\ref{eq-BDstar}), extrapolated to zero recoil, is currently know from experiment to a precision of $2.5\%$~\cite{HFAG} with a 2020 BelleII projection of $2\%$ for $B\to D^*\tau\nu$~\cite{BelleII-future}.
These precisions translate to target uncertainties on the lattice calculation of $\mathcal{F}(1)$ of $1.3\%$ now and $1\%$ in 2020.
Fig.~\ref{fig-G1F1} compares lattice calculations of $\mathcal{F}(1)$ to these target uncertainties.

The decays $B_{(s)}\to D_{(s)} l\nu$ allow alternative determinations of $|V_{cb}|$.
For $B^\pm$ decay, the differential decay rate is given by
\begin{equation}
\frac{d\Gamma (B^\pm\to D^0 l\nu)}{dw} \frac{48\pi^3}{G_F^2 |\eta_{EW}|^2 M_D^3 (M_B+M_{D})^2 (w^2-1)^{3/2}} = |V_{cb}|^2 |\mathcal{G}(w)|^2\ +\  \mathcal{O}\left(\frac{m_ l^2}{q^2}\right).
\label{eq-BD}
\end{equation}
Experiment has determined the left-hand side to $7.2\%$ and BelleII projections indicate a $3\%$ measurement for $B\to D\tau\nu$ by 2020.\footnote{Belle is expected to report an updated measurement of $B\to D^{(*)}\tau\nu$ in the coming months.  The $B_s\to D_s l\nu$ decay has not yet been observed.}  The equivalent target precisions for $|\mathcal{G}(1)|$, $3.6\%$ and $1.5\%$, are plotted in Fig.~\ref{fig-G1F1} along with existing lattice calculations.
As with $B\to D^*l\nu$ most work to date has focused on zero recoil.

FNAL/MILC published results~\cite{FNAL-Vcb} for $B\to D^*l\nu$ at zero recoil, obtaining an error commensurate with the current experimental precision.
Using the MILC $N_f=2+1$ asqtad ensembles with asqtad light and FNAL charm and bottom valence quarks, they simulate at five lattice spacings (0.045, 0.06, 0.09, 0.12, and 0.15 fm) and with pions as light as 174 MeV, and
found their leading error to be heavy-quark discretization effects.
In an effort to reduce heavy-quark discretization effects, FNAL/MILC and SWME are employing the Oktay-Kronfeld action~\cite{OK} and have found improvements in the $B$ meson dispersion relation and hyperfine splitting~\cite{Jang}.
In a related effort, SWME~\cite{Bailey} plans to calculate $B\to D^* l\nu$ at zero recoil using the MILC $N_f=2+1+1$ HISQ gauge fields, physical light quark masses, HISQ light and charm and Oktay-Kronfeld bottom valence quarks, and an improved heavy-light current with on-shell improvement through $\mathcal{O}(p^3)$.
HPQCD~\cite{Na:LAT14} showed preliminary results for $B_{(s)}\to D_{(s)} l \nu$ from a calculation using 
the MILC $N_f=2+1$ asqtad gauge fields with HISQ light and charm and NRQCD bottom valence quarks,
two lattice spacings (0.09 and 0.12 fm),
and pions as light as 260 MeV.
This work includes the kinematic dependence of the form factors and branching fractions for $B\to Dl\nu$ and $B_s\to D_s l\nu$, ratios of which are useful in the analysis of $\mathcal{B}(B_s\to\mu\mu)$ (see, e.g., Ref.~\cite{FNAL:BDBsDs}).
Since the conference, FNAL/MILC has reported results for $B\to D l \nu$ form factors near zero recoil~\cite{FNAL:BD}.

\begin{figure}[t!]
{\scalebox{0.9}{\includegraphics[angle=0,width=0.5\textwidth]{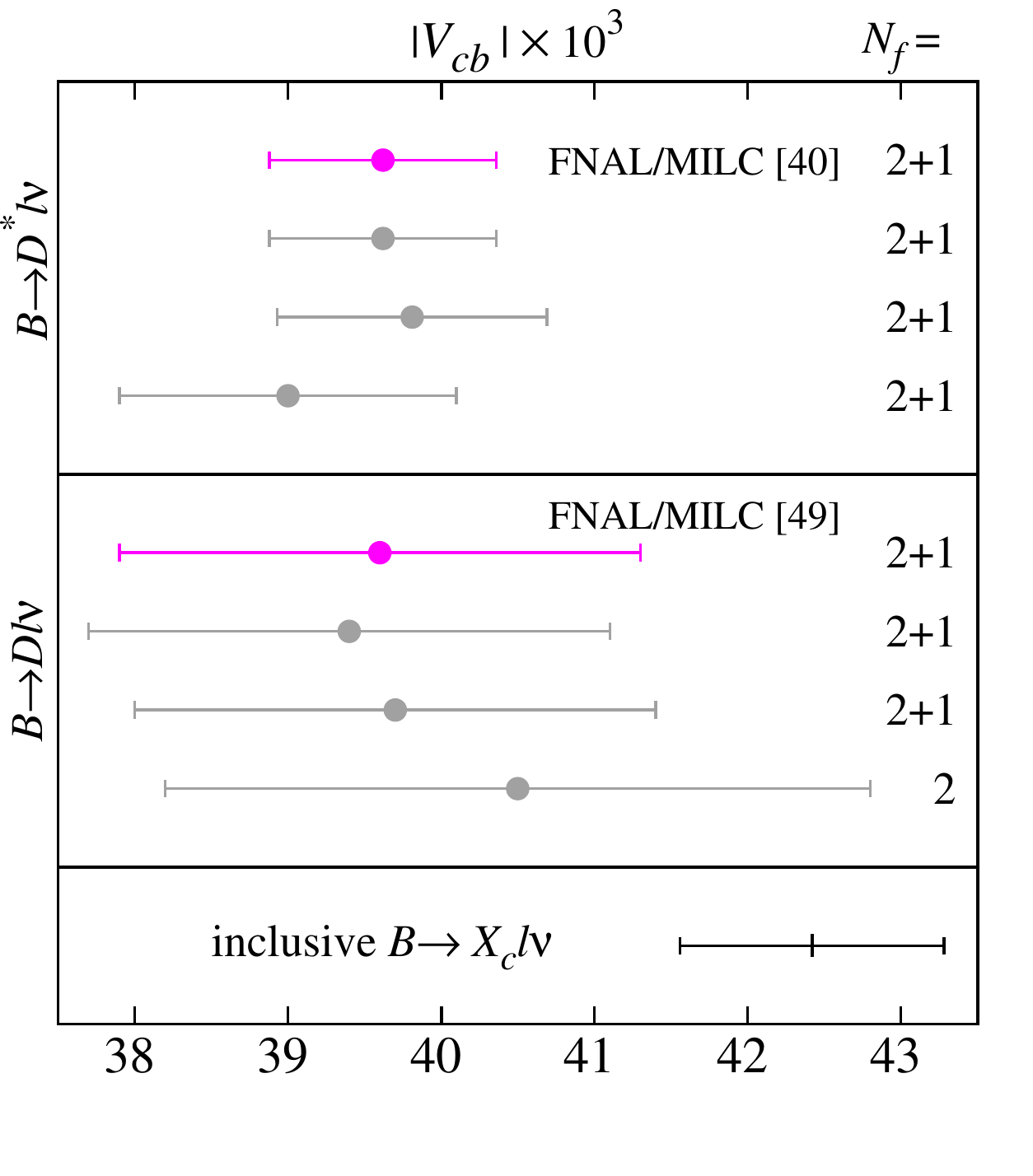}}}
\hspace{0.0in}
{\scalebox{1.0}{\includegraphics[angle=0,width=0.5\textwidth,bb=30 -20 900 500]{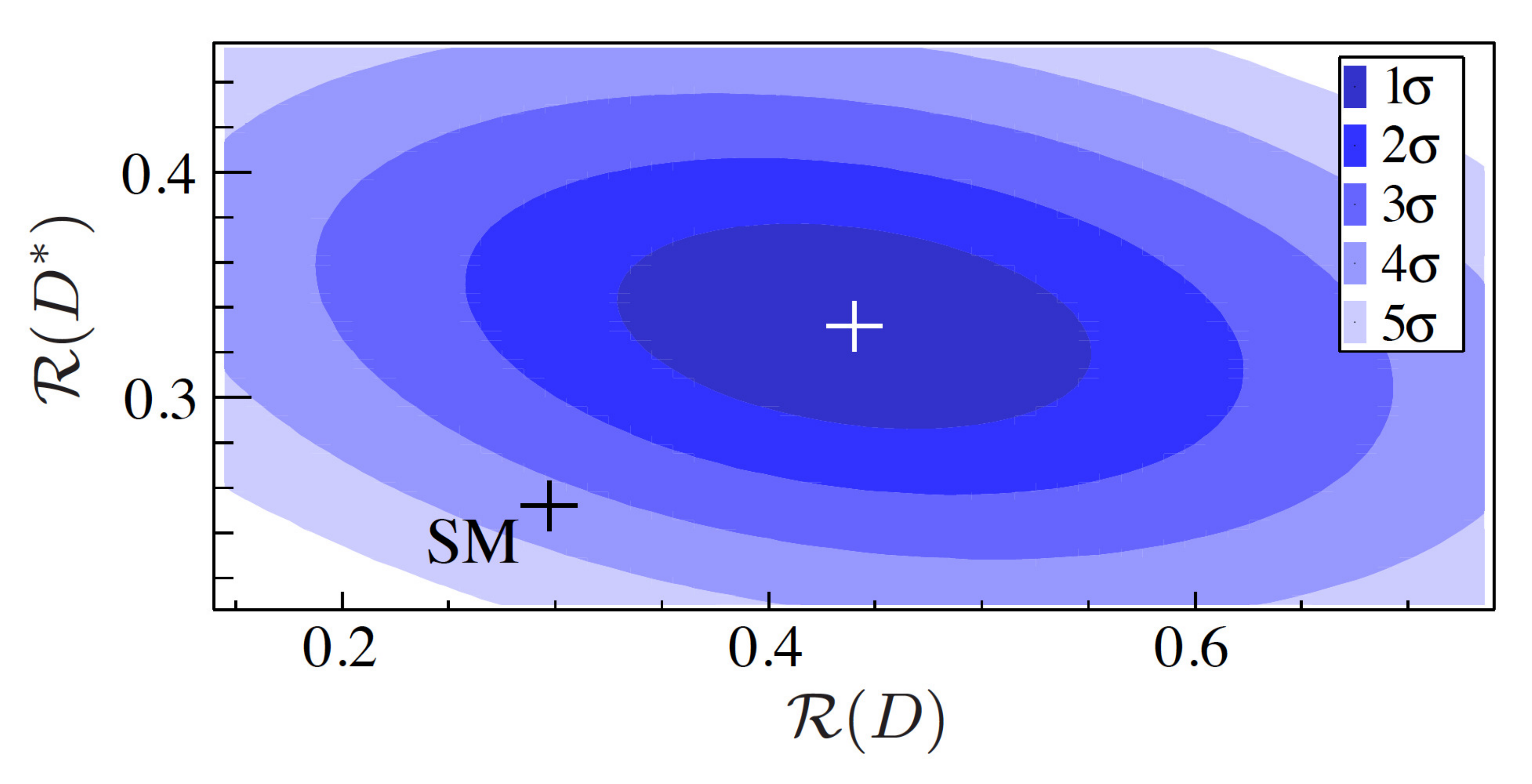}}}
\caption{(left) Comparison of $|V_{cb}|$ values obtained using recent results from FNAL/MILC~\cite{FNAL-Vcb, FNAL:BD}, shown in magenta, and previous year's results in gray~\cite{FLAG}.  (right) BABAR has measured a discrepancy of just over $3\sigma$ in the correlated ratios of branching fractions~\cite{BABAR_R} defined in Eq.~(\protect\ref{eq-RD}).}
\label{fig-Vcb_R}
\end{figure}

Fig.~\ref{fig-Vcb_R} compares values of $|V_{cb}|$ obtained from lattice calculations of the exclusive semileptonic decays $B\to D^{(*)}l\nu$ and from the inclusive decay $B\to X_c l\nu$~\cite{Vcb_incl}, revealing a persistent $3\sigma$ tension.  
Also show in Fig.~\ref{fig-Vcb_R}, a $3\sigma$ tension between experiment and the SM has been measured by BABAR~\cite{BABAR_R} in the ratios 
\begin{equation}
\mathcal{R}(D^{(*)}) = \frac{\mathcal{B}(B\to D^{(*)}\tau\nu)}{\mathcal{B}(B\to D^{(*)}\mu\nu)}.
\label{eq-RD}
\end{equation}
In light of these tensions it is important to improve upon lattice determinations of the $B\to D^{(*)}l\nu$ form factors to fully leverage the precision of experimental results.  
To this end, improved determinations of form factor shapes and measured rates reported as functions of $q^2$, or other suitable kinematic variable, would be particularly useful.

\section{Rare Processes}
\label{sec-rare}
Rare processes involve FCNCs and only occur in the SM via vacuum fluctuations.  
This introduces a loop suppression factor, often accompanied by a combination of the GIM mechanism and further Cabibbo suppression.
This SM suppression makes such processes potentially susceptible to new physics effects and provides a strong motivation for their study.

\subsection{Rare $B_{(s)}$ Decays}
\label{sec-decays}
\begin{figure}
\centering
\includegraphics[width=0.329\textwidth]{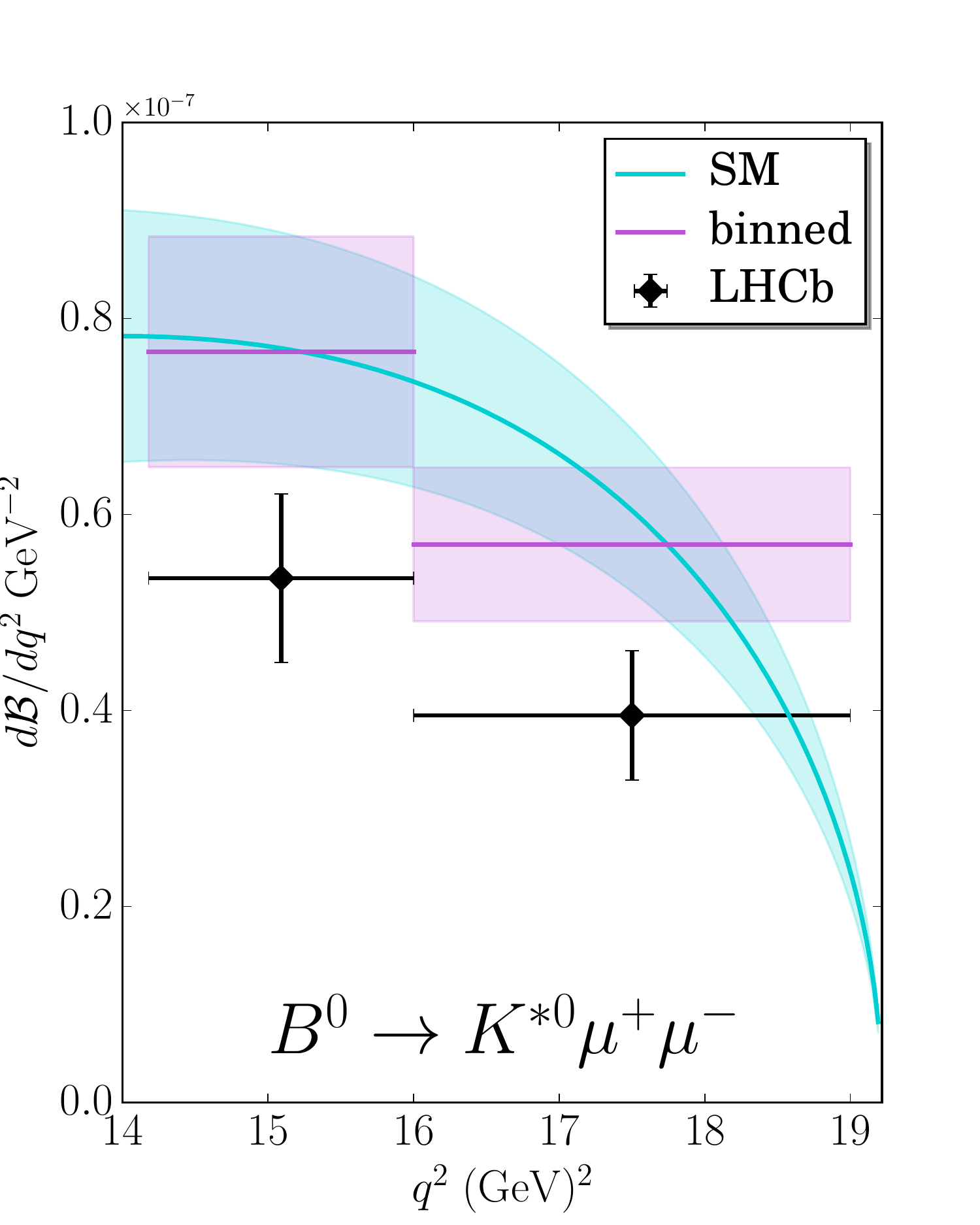}
\includegraphics[width=0.329\textwidth]{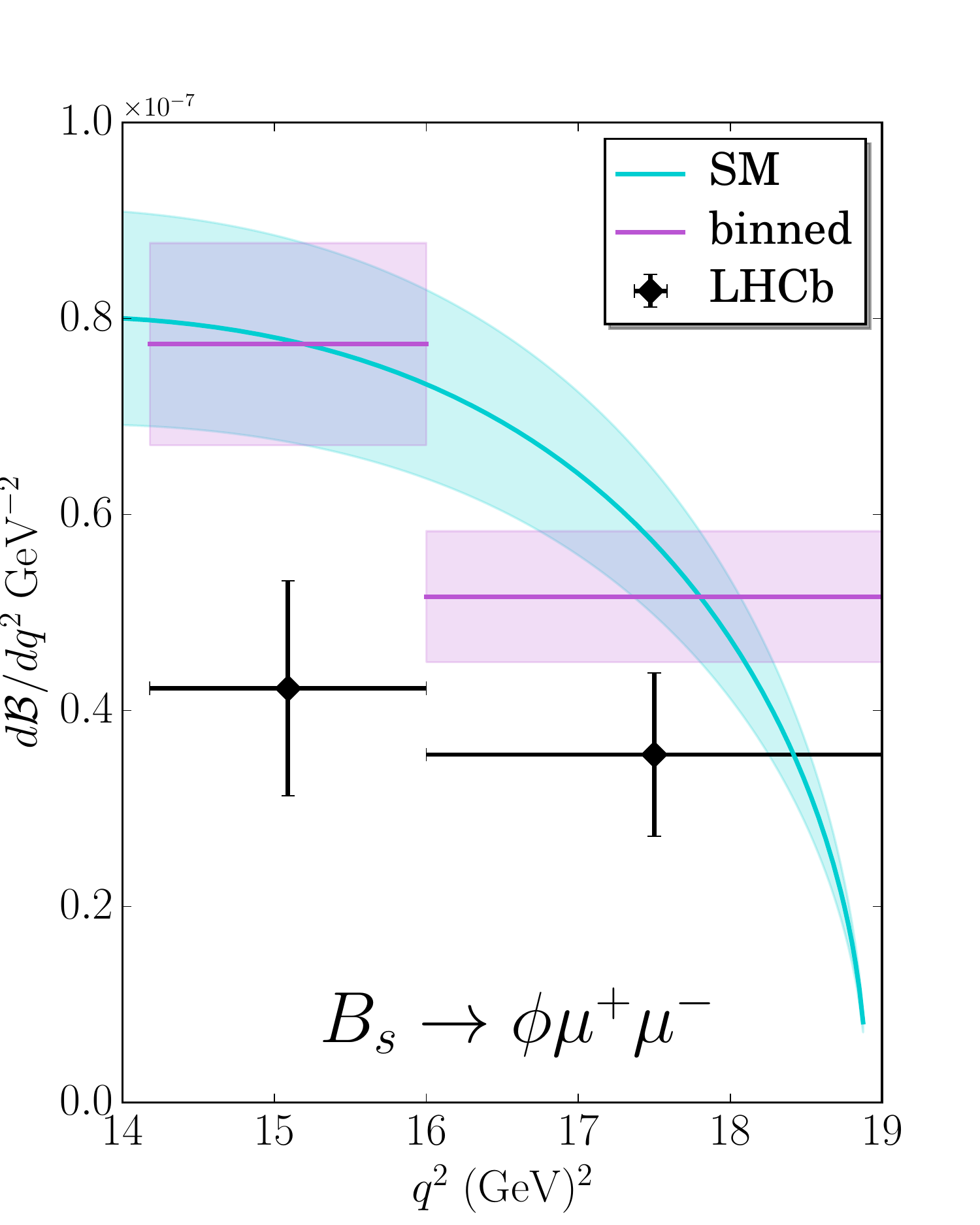}
\includegraphics[width=0.329\textwidth]{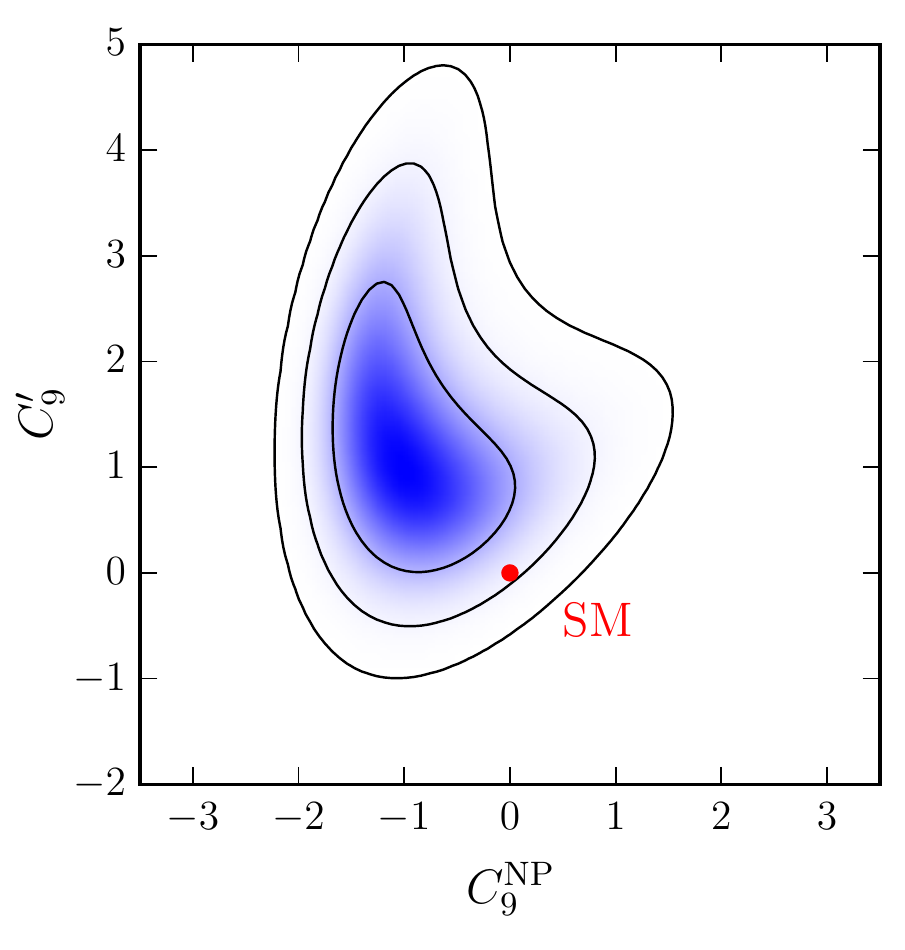}
\caption{\label{fig-Wingate} From Ref.~\cite{Horgan:2015}:  
the differential branching fractions for 
(left) $B^0\to K^{*0}\mu^+\mu^-$, 
(middle) \mbox{$B_s\to \phi\mu^+\mu^-$}, and 
(right) the resulting constraint on the Wilson coefficients $C'_9$ and $C_9^{\rm NP}$, where \mbox{$C_9 = C_9^{\rm SM} + C_9^{\rm NP}$}, obtained from combination with experiment.  
The SM value lies just inside the $2\sigma$ contour.}
\end{figure}
In rare decays, as with the tree-level decays discussed above, the operator product expansion leads to a factorization of short distance (high energy) physics and long distance (low energy) physics.
The model-dependent short distance physics responsible for the flavor-changing interactions occurs at a scale of $M_W$ (or heavier) in the SM (models of new physics) and is characterized by the Wilson coefficients $C_i^{(}{'}^{)}$.
Long distance physics associated with the hadronization of observed asymptotic states is contained in hadronic matrix elements of local operators.
In FCNC $b\to s$ decays this is evident in the effective Hamiltonian density,
\begin{equation}
\mathcal{H}_{\rm eff}^{b\to s} = - \frac{4 G_F}{\sqrt{2}} V_{tb} V_{ts}^* \sum_i ( C_i O_i + C'_i O'_i),
\end{equation}
with local operators given by, e.g.,
\begin{equation}
O_7^{(}{\!'}^{)} = \frac{em_b}{16\pi^2}\ \bar s \sigma_{\mu\nu} P_{R(L)} b\ F^{\mu\nu}, \hspace{0.08in}
O_9^{(}{\!'}^{)} = \frac{e^2}{16\pi^2}\ \bar s \gamma_\mu P_{L(R)} b\ \bar l \gamma^\mu l,\ {\rm and} \hspace{0.08in}
O_{10}^{(}{\!\!\!\!'}^{)} = \frac{e^2}{16\pi^2}\ \bar s \gamma_\mu P_{L(R)} b\ \bar l \gamma^\mu\gamma_5 l.
\end{equation}
Observation, coupled with lattice calculation of the matrix elements, constrains the $C_i^{(}{'}^{)}$ and provides evidence in support of, or against, the model responsible for the flavor-changing interactions.

The authors of Ref.~\cite{Horgan:2015} calculate form factors for the decays $B\to K^*ll$, $B_s\to\phi ll$, and $B_s\to~\!\!K^*ll$.
The daughter vector mesons in these decays are unstable.
Though associated threshold effects are not accounted for in current analyses, recent developments~\cite{Raul} suggest a path for their inclusion.
The calculations of Ref.~\cite{Horgan:2015} use the MILC $N_f=2+1$ asqtad ensembles at two lattice spacings (0.09 and 0.12 fm),
asqtad light and strange and NRQCD bottom valence quarks,
and pion masses down to 313 MeV.
Fig.~\ref{fig-Wingate} reveals a slight tension between the SM and experiment, and shows the resultant constraints obtained on new physics Wilson coefficients.
FNAL/MILC is calculating form factors for $B\to Kll$ decays using the MILC $N_f=2+1$ asqtad ensembles 
at four lattice spacings (0.045, 0.06, 0.09, and 0.12 fm),
with asqtad light and strange and FNAL bottom valence quarks, 
and pions as light as 174~MeV.
As extensions of $B\to\pi l\nu$ calculations discussed above, FNAL/MILC~\cite{Du} and HPQCD~\cite{Bouchard:BPi} are also calculating the form factors for $B\to\pi ll$.
All FCNC rare decays are subject to potential non-factorizable effects associated with $c\bar c$ resonances, see Ref.~\cite{Horgan:2015} for discussion.
It is not clear at present how best to deal with these effects.

\subsection{Neutral Meson Mixing}
\label{sec-mixing}
Expanding the off-diagonal term of the neutral meson $\mathcal{M}^0$ mass matrix to second order in the weak interaction gives
\begin{equation}
M_{12} - \frac{i}{2} \Gamma_{12} = \sum_{X; jk} \frac{ C_j^{\Delta_f=1} C_k^{\Delta_f=1} \langle \bar\mathcal{M}^0 | O_j^{\Delta_f=1} | X\rangle \langle X | O_k^{\Delta_f=1} | \mathcal{M}^0 \rangle }{M_{\mathcal{M}^0} - E_X  +i\epsilon} + \sum_i C_i^{\Delta_f=2} \langle \bar\mathcal{M}^0 | O_i^{\Delta_f=2} | \mathcal{M}^0\rangle,
\end{equation}
where $\Delta_f$ gives the change in flavor quantum number.
The term with two $\Delta_f=1$ interactions is non-local for (near) onshell intermediate state $|X\rangle$.
The calculation of these long distance effects is notoriously difficult, though recent progress has been made in kaon mixing~\cite{Sachradja}.
The short distance term with a single $\Delta_f=2$ interaction is local and amenable to calculation.
The space of all possible hadronic matrix elements of dimension 6 effective four quark operators is spanned by the basis
\begin{eqnarray}
O_1^{\Delta_f=2}\!\!&=&\!\bar Q^\alpha \gamma_\mu L q^\alpha\  \bar Q^\beta \gamma_\mu L q^\beta,  \ \ \ O_2^{\Delta_f=2}=\bar Q^\alpha L q^\alpha\  \bar Q^\beta L q^\beta,\ \ \  O_3^{\Delta_f=2}=\bar Q^\alpha L q^\beta\  \bar Q^\beta L q^\alpha,  \nonumber\\
&& O_4^{\Delta_f=2}=\bar Q^\alpha L q^\alpha\  \bar Q^\beta R q^\beta,\ \ \ {\rm and}\ \ O_5^{\Delta_f=2}=\bar Q^\alpha L q^\beta\  \bar Q^\beta R q^\alpha,
\label{eq-4qops}
\end{eqnarray}
where $L/R = (1\pm\gamma_5)/2$ and color indices are explicit.
Mixing matrix elements are historically parameterized in terms of bag parameters $B^{(i)}_{\mathcal{M}^0}$\,, e.g.,
\begin{equation}
\langle \bar\mathcal{M}^0 | O_1^{\Delta_f=2} | \mathcal{M}^0\rangle = \frac{2}{3} f^2_{\mathcal{M}^0}\, M^2_{\mathcal{M}^0}\, B^{(1)}_{\mathcal{M}^0}.
\end{equation}

\subsubsection{$D$ mixing}
\label{sec-Dmix}
In SM $D$ mixing, the short distance contribution is doubly Cabibbo suppressed (by a factor of $|V_{ub}V_{cb}^*|^2$) and the long distance contribution dominates.  
In the long distance mixing of the SM, the lighter down and strange quarks of the first two generations dominate and CP violation is negligible.
As a result, strong constraints can be placed on CP violating observables associated with new physics, generally associated with short distance interactions described by the local effective four quark operators of Eq.~(\ref{eq-4qops}).
The lattice calculation of hadronic matrix elements of these operators is therefore phenomenologically important~\cite{Dpheno}.

ETM has recently published the first unquenched calculation of the hadronic contribution to short-distance $D$ mixing~\cite{ETM:Dmix}.
Using the ETM $N_f=2$ configurations at four lattice spacings (0.05, 0.07, 0.09, and 0.10 fm), 
$\mathcal{O}(a)$ improved Ostwerwalder-Seiler valence quarks,
and pions as light as 280 MeV,
they calculate the bag parameters to $3-5\%$ precision.
This is about five time more precise than current equivalent experimental precision~\cite{UTfit:Dmix, LHCb:Dmix} and on par with 2020 expectations~\cite{Dmix:future}.
FNAL/MILC~\cite{Jason} presented preliminary results for the mixing matrix elements from a calculation using the MILC $N_f=2+1$ asqtad configurations at four lattice spacings (0.045, 0.06, 0.09, and 0.12 fm),
asqtad light and FNAL charm valence quarks,
and with pions as light as 177 MeV.

\subsubsection{$B_{(s)}$ mixing}
\label{sec-Bmix}
Because $B_{(s)}$ mixing is dominated, in the SM and beyond, by short distance contributions, there are phenomenologically relevant SM quantities to be calculated in addition to the hadronic matrix elements (bag parameters) discussed above.
These quantities include the SM oscillation frequency $\Delta M_{(s)}$ and the SU(3) breaking ratio $\xi$, closely related to the ratio of CKM matrix elements $|V_{td}/V_{ts}|$,
\begin{equation}
\Delta M_{(s)} = C_1^{\Delta_b=2} \langle \bar B^0_{(s)} | O_1^{\Delta_b=2} | B^0_{(s)}\rangle \qquad{\rm and}\qquad \xi= f_{B_s} \sqrt{B^{(1)}_{B_s}}\ /\ f_B \sqrt{B^{(1)}_B} .
\end{equation}
$B$ mixing oscillation frequencies have been measured to sub-percent precisions~\cite{PDG}.

In addition to the $B_{(s)}$ decay constants discussed above, in Ref.~\cite{Aoki} Aoki~{\it et al.} also calculate the hadronic contribution to SM $B_{(s)}$ mixing.
In Ref.~\cite{Christine} HPQCD gives a status report on the first lattice calculation of $B_{(s)}$ meson mixing parameters with physical light quark masses.  
This calculation uses three lattice spacings (0.09, 0.12, and 0.15 fm) from the MILC $N_f=2+1+1$ HISQ ensembles with radiatively-improved NRQCD $b$ and HISQ light and strange valence quarks.
Early results show impressive precision for SM mixing parameters.
FNAL/MILC updated~\cite{Aida} their nearly complete calculation of mixing parameters relevant in the SM and beyond.  This calculation uses four lattice spacings (0.045, 0.06, 0.09, and 0.12 fm) of the MILC $N_f=2+1$ asqtad ensembles, asqtad light and strange and FNAL $b$ valence quarks, with pion masses as light as 177 MeV.
Errors of $\sim 9\%$ are expected for the matrix element $\langle \bar B^0_{(s)} | O_1^{\Delta_b=2} | B^0_{(s)} \rangle$, 
$10-15\%$ for remaining matrix elements,
and better than $2\%$ for $\xi$.

\section{Summary}
\label{sec-summary}
The past year has seen significant activity in lattice heavy flavor physics.
By my count there are 35 recent or ongoing calculations with 20 talks or posters presented at this conference.
Besides conveying the level of lattice activity over the past year, this review compared how we are doing relative to current and projected experimental precision.
We are, generally, poised to fully leverage expected experimental results, with the possible exception of $D$ semileptonic decays where several works underway should improve the situation.
In addition to the determination of CKM matrix elements, decay constants and form factors are also used to parameterize hadronic contributions to many phenomenological observables, making improved calculations important independent of the experimental results discussed in this review.
In light of recent lattice results, we reviewed several tensions between nature and the SM, including: 
leptonic and semileptonic determinations of $|V_{cs}|$,
inclusive versus exclusive determinations of $|V_{ub}|$ and $|V_{cb}|$,
the correlated ratios $\mathcal{R}(D)$ and $\mathcal{R}(D^*)$, and
branching fractions for some rare $B_{(s)}$ decays.
Continued coordination with phenomenologists and experimentalists is crucial to extracting as much as possible from each measurement and calculation, and to identifying useful new observables in our search for cracks in the SM.

\section*{Acknowledgements}
I thank the organizers of Lattice 2014; first, for the invitation to give this review and second, for a well-organized conference on the beautiful campus of Columbia University in bustling Manhattan.
Thanks also to the long list of colleagues who shared information regarding their ongoing calculations and allowed me to discuss their work with you.


\end{document}